\documentclass[journalca]{IEEEtran}
\pdfoutput=1

\usepackage[bookmarks=false]{hyperref}
\hypersetup{
  colorlinks,
  linkcolor={red!50!black},
  citecolor={blue!50!black},
  urlcolor={blue!80!black}
}

\usepackage{color}
\usepackage{amsmath,gensymb}
\usepackage{amssymb}
\usepackage{stmaryrd}
\usepackage{todonotes}
\usepackage{subfig}
\usepackage{epstopdf}
\usepackage{tikz}
\usepackage[noadjust]{cite}

\title{The Unfitted Discontinuous Galerkin Method for Solving the EEG Forward Problem}

\author{Andreas~N\"u\ss ing, 
     Carsten~H.~Wolters, 
      Heinrich~Brinck,
      Christian~Engwer
\thanks{Andreas~N\"u\ss ing is with the Institute for Bioinformatics and Chemoinformatics, Westf{\"a}lische Hochschule, Recklinghausen, Germany, with the Institute for Computational and Applied Mathematics, University of M{\"u}nster, Germany and with the Institute for Biomagnetism and Biosignalanalysis, University of M{\"u}nster, Germany.}
\thanks{C.\ H.\ Wolters is with the Institute for Biomagnetism and Biosignalanalysis, University of M{\"u}nster, Germany.}
\thanks{H.\ Brinck is with the Institute for Bioinformatics and Chemoinformatics, Westf{\"a}lische Hochschule, Recklinghausen, Germany.}
\thanks{Christian~Engwer is with the Institute for Computational and Applied Mathematics and the Cluster of Excellence EXC 1003, Cells in Motion, CiM, University of M{\"u}nster, Germany.}
}

\newcommand{\changedReview}[1]{\textcolor{black}{#1}}
\newcommand{\NN}{\mathbb{N}}
\newcommand{\PP}{\mathbb{P}}
\newcommand{\QQ}{\mathbb{Q}}
\newcommand{\RR}{\mathbb{R}}
\newcommand{\Dd}{\mathcal{D}}
\newcommand{\Ii}{\mathcal{I}}
\newcommand{\Pp}{\mathcal{P}}

\newcommand{\Tt}{\mathcal{T}}
\newcommand{\rdmp}{\operatorname{RDM\%}}
\newcommand{\magp}{\operatorname{MAG\%}}

\newcommand{\Uana}{U_{\text{ana}}}
\newcommand{\Unum}{U_{\text{num}}}

\begin{document}
\maketitle
\begin{abstract}~
  \emph{Objective:}
  The purpose of this study is to introduce and evaluate the unfitted discontinuous Galerkin finite element method (UDG-FEM) for solving the electroencephalography (EEG) forward problem.
  \emph{Methods:}
  This new approach for source analysis does not use a geometry conforming volume triangulation, but instead uses a structured mesh that does not resolve the geometry.
  The geometry is described using level set functions and is incorporated implicitly in its mathematical formulation.
  As no triangulation is necessary, the complexity of a simulation pipeline and the need for manual interaction for patient specific simulations can be reduced and is comparable with that of the FEM for hexahedral meshes. 
  In addition\changedReview{,} it maintains conservation laws on a discrete level.
  Here, we present the theory for UDG-FEM forward modeling, its \changedReview{verification} using quasi\changedReview{-}analytical solutions in multi-layer sphere models and an evaluation in a comparison with a discontinuous Galerkin (DG-FEM) method on hexahedral and on conforming tetrahedral meshes.
  We furthermore apply the UDG-FEM forward approach in a realistic head model simulation study.
  \emph{Results:}
  The given results show convergence and indicate a good overall accuracy of the UDG-FEM approach. 
  UDG-FEM performs comparable or even better than DG-FEM on a conforming tetrahedral mesh while providing a less complex simulation pipeline. When compared to DG-FEM on hexahedral meshes, an overall better accuracy is achieved.
  \emph{Conclusion:}
  The UDG-FEM approach is an accurate, flexible and promising method to solve the EEG forward problem.
  \emph{Significance:}
  This study shows the first application of the UDG-FEM approach to the EEG forward problem.\\
\end{abstract}
\section{Introduction}
The localization of current sources in the human brain from surface electroencephalography (EEG) measurements (the
inverse problem) requires a model for the forward problem, i.e., the determination of surface potentials from current sources in
the cortical sheet of the human brain \cite{Haemaelaeinen1993,Mun2012}.
Several different approaches have been proposed to solve the EEG forward problem.
When approximating the head by a multi-compartment sphere model, quasi\changedReview{-}analytical series expansion formulas are available \cite{Munck1993}.
More realistic methods are based on head volume conductor representations obtained from quasi-noninvasive magnetic resonance imaging (MRI).
Under these approaches are boundary element methods (BEM) \cite{Mosher1999,Acar2010,Gramfort2011,Stenroos2012}, 
finite volume methods (FVM) \cite{Cook2006}, finite difference methods (FDM) \cite{CHW:Wen2008,Vatta2008,CHW:Mon2014} and 
-finite element methods (FEM) \cite{CHW:Wei2000,Schimpf2002,CHW:Gen2004,Wolters2007c,Lew2009b,Vallaghe2010,Pursiainen2011,Vorwerk2012,CHW:Med2015}.
\changedReview{The method presented in this paper is closely related to FEM approaches.
  The latter were shown to produce highly accurate solutions \cite{Vorwerk2012}.
  In combination with transfer matrices and efficient linear solvers \cite{CHW:Wei2000,CHW:Gen2004,Mun2012}, the computational effort could be reduced.
  By using volumetric meshes, they are able to handle anisotropic conductivities and complex model geometries \cite{Vorwerk2014}.}

A common FEM approach is to use constrained Delaunay tetrahedralization (CDT) to generate a conforming tetrahedral mesh which is 
constructed from non-intersecting surface triangulations \cite{Vorwerk2014}. 
Although the method allows smooth tissue surface representations, the process of constructing proper non-intersecting surfaces might become an involved task.
\changedReview{In order to allow CDT meshing, unrealistic model features might be introduced, such as the artificial closing of holes (e.g. optical canals or foramen magnum).
Furthermore, the surfaces are required to be nested, while they are touching in a realistic scenario (e.g. the inner skull surface and the brain surface).}

\changedReview{Such limitations can be circumvented by using hexahedral models.
  They can be directly generated from voxel-based MRI images and are used in several source analysis applications \cite{Schimpf2002, CHW:Par2015,CHW:Ayd2015}.
 The hexahedral approach keeps the advantages of the FEM mentioned above, while reducing the effort for creating individual head models.}

However, one problem of FEM modeling using regular hexahedral meshes is the possibility of the occurrence of skull leakages: In areas where the skull is very thin, for example the temporal bone, where skull thickness is 2 mm or even less \cite[Table 2]{Kwon2006}, physically 
unsound leakages of current through the skull might occur in regular hexahedral approaches with insufficient resolution \cite{Sonntag2013,Engwer2015}. It was shown in \cite{Dannhauer2011} that  
an appropriate skull modeling is of special importance for accurate EEG forward modeling. The effects of skull leakages
might thus lead to especially significant errors when hexahedral FEM modeling is used in combination with insufficient resolution. 
A way to alleviate this problem has been proposed in \cite{Engwer2015}, where
instead of the node based approach of the Lagrange or Continuous Galerkin FEM (CG-FEM), 
a cell based Discontinuous Galerkin FEM (DG-FEM) method was introduced. In this paper, we will 
therefore focus on DG-FEM approaches.

A second problem of regular hexahedral FEM approaches is the stair-case like representation of smooth head tissue surfaces, producing geometrical inaccuracies that lead to modeling errors, especially in combination with lower resolutions. Therefore, methods have been developed to reduce such geometrical errors.
In \cite{Wolters2007}, the structured mesh was modified to obtain a smoother interface between different tissue compartments.
However, this method might still suffer from skull leakages and geometry adaptation is limited by the need for positive Jacobian determinants.
In \cite{Vallaghe2010}\changedReview{,} a so-called immersed FEM has been proposed for the EEG forward problem.
It combines the simplicity of a hexahedral CG-FEM approach with the accuracy of a conforming tetrahedral approach 
with regard to the modeling of smooth tissue surfaces. Using a level set function to represent the surface, the local 
basis functions of the hexahedral approach were modified to conform to the tissue compartments. 
It was shown that the accuracy of a conforming tetrahedral approach could be met, while offering a simpler simulation pipeline. 

In this paper, we present the unfitted discontinuous Galerkin finite element method (UDG-FEM) to solve the EEG forward problem.
The method was first introduced in the context of micro-scale simulations in porous media \cite{Bastian2009} and has since then 
found different applications \cite{Heimann2013, Engwer2014, Burman2015}.
It combines the advantages of the DG-FEM approach on a structured hexahedral mesh with 
implicit surface representations given by level set functions and thereby the avoidance of skull leakages and 
inappropriate modeling of smooth tissue surfaces. As we will show, it outperforms a state-of-the-art 
DG-FEM approach and leads to simpler forward modeling pipelines because it is not restricted
to nested compartments and can handle intersecting surfaces such as touching surfaces of inner skull and brain.

Our paper is structured as follows: After a thorough description of the theory in section \ref{sec:theory}, we provide methodological aspects of our \changedReview{verification} and evaluation in section \ref{sec:methods}. Results are then presented in section \ref{sec:results} and 
discussed in section \ref{sec:discussion}. The paper ends with concluding remarks in chapter \ref{sec:conclusion}.

\section{Theory}
\label{sec:theory}
\subsection{A discontinuous Galerkin (DG-FEM) method for solving the EEG forward problem}
The EEG forward problem can be solved by providing a solution to Poisson's equation which results from the quasi\changedReview{-}static Maxwell's equations \cite{Haemaelaeinen1993,Mun2012}.
Let $\Omega\subset\RR^3$ denote the head domain and $\partial\Omega$ its surface.
The task is then to find \changedReview{the electric scalar potential $u$} for which
\begin{alignat}{2}
  \nabla\cdot\sigma\nabla u&=f&\quad&\text{in }\Omega\label{eq:poisson} \\
  \sigma\nabla u\cdot n&=0&\quad&\text{in }\partial\Omega\label{eq:poisson_boundary}
\end{alignat}
holds.
\changedReview{$\sigma:\Omega\to\operatorname{Sym_3}(\RR)$} denotes the tissue \changedReview{conductivity tensor of second rank} and $f$ the \changedReview{source current density}.
For each $x\in\Omega$, $\sigma(x)$ is assumed to be symmetric and positive definite, but it is allowed to be 
discontinuous over tissue boundaries.
A common source model, which we will also use here, is the mathematical dipole.
It is represented by a position $x_0\in\Omega$ and a moment vector $M\in \RR^3$ and given as:
\begin{equation}
  f(x) = M\cdot\nabla\delta(\changedReview{x-x_0})\label{eq:dipole}
\end{equation}
where $\delta$ denotes Dirac's delta distribution.
Note that the mathematical dipole model is not a function and that its divergence has to be considered in distributional sense.
\changedReview{Note also that for \eqref{eq:poisson} to hold in a strong sense, more rigorous regularity assumptions on $u$, $\sigma$ and $f$ would be required, so that \eqref{eq:poisson} should be seen in a symbolic sense.}

In \cite{Engwer2015}, a discontinuous Galerkin (DG-FEM) formulation for \eqref{eq:poisson} has been derived, which we will briefly recall here.
\changedReview{For a detailed description and an analysis of the properties of DG-FEM, we refer to \cite{Engwer2015}.}
\changedReview{Note that this formulation also serves as the foundation of the unfitted discontinuous Galerkin method (UDG-FEM) presented below.}
First, we introduce a volume triangulation of $\Omega$:
\begin{align}
\Tt_h=\lbrace E_i | i\in\Ii=\lbrace 0,\dots,N-1\rbrace\rbrace\\
E_i\cap E_j =\changedReview{\varnothing}~\forall i\not=j,\quad\bigcup_{i\in\Ii}\overline{E}_i=\overline{\Omega}
\end{align}
of open sets $E_i\subset\RR^3$. \changedReview{For such an open set $E$, $\overline{E}$ denotes its closure, i.e. $\overline{E}=E\cup\partial E$.}
\changedReview{The mesh width $h\in\RR$ is defined as} $h:=\max\lbrace \operatorname{diam}(E): E\in\Tt_h\rbrace$.
In the following, we will restrict the description to the case that all elements are either tetrahedrons or hexahedrons.
The skeleton $\Gamma_h$ of $\Tt_h$ is defined as $\Gamma_h:=\lbrace \gamma_{i,j}=\overline{E}_i\cap\overline{E}_j: E_i,E_j\in\Tt_h, i\not=j, |\gamma_{i,j}|>0\rbrace$.
Let $V_h^k=\lbrace u\in L_2(\Omega) : u|_E\in \Pp^k(E)\ \forall E\in\Tt_h\rbrace$ denote the broken polynomial space on $\Tt_h$.
$\Pp^k(E)$ denotes a space of polynomials on $E$ of degree $k\in\NN$.
In the following, we will assume that $\sigma$ is constant on each $E_i$ and denote its value by $\sigma_i$.
The discontinuous Galerkin method for solving \eqref{eq:poisson}, which we will use, then reads:
Find $u_h\in V_h^k$ such that \changedReview{for all test functions $v_h\in V_h^k$}
\begin{align}
  a(u_h,v_h)+J(u_h,v_h)&= l(v_h)
  \label{eq:dg}
\end{align}
holds. The bilinear forms $a$ and $J$ are given as\changedReview{
\begin{align}
  a(u_h,v_h) =& \int_{\Omega} \sigma\nabla u_h\cdot\nabla v_h dx -\int_{\Gamma_h}\llbracket u_h\rrbracket\cdot\langle\sigma\nabla v_h\rangle ds\\
&-\int_{\Gamma_h} \llbracket v_h\rrbracket\cdot\langle\sigma\nabla u_h\rangle ds\\
  J(u_h,v_h) =& \eta\int_{\Gamma_h} \frac{\tau_\gamma}{h_\gamma}\llbracket u_h\rrbracket\cdot\llbracket v_h\rrbracket ds\\
  l(v_h) =& \int_\Omega fv_hdx
  \label{eq:dg_forms}
\end{align}}
The jump $\llbracket u_h\rrbracket\changedReview{\in\RR^3}$ on the intersection between two elements $E_i$ and $E_j$ with unit outer normals $n_i\changedReview{\in\RR^3}$ and $n_j\changedReview{\in\RR^3}$, respectively, is defined as $\llbracket u_h\rrbracket := u_h|_{E_i}n_i+u_h|_{E_j}n_j$.
\changedReview{The weighted average of the flux of $u_h$ on the interface is denoted by $\langle\sigma\nabla u_h\rangle\in\RR^3$.}
With $\delta_i := n_i^t\sigma_in_i$ and $\delta_j :=  n_j^t\sigma_jn_j$, this can be defined as\changedReview{
\begin{align}
  \langle \sigma\nabla u_h\rangle := \frac{\delta_i}{\delta_i+\delta_j}\sigma_i\nabla u_h|_{E_i}
                                   + \frac{\delta_j}{\delta_i+\delta_j}\sigma_j\nabla u_h|_{E_j}
\end{align}}
\changedReview{Note that both the jump of the potential and the weighted average of the flux are vector valued quantities and that the normals $n_i$ and $n_j$ are opposing vectors, i.e. $n_i=-n_j$.}
The factor $\tau_\gamma$ scales the penalty term \changedReview{$J$} at conductivity jumps on an edge $\gamma=\gamma_{i,j}$.
It is defined as the harmonic average of $\delta_i$ and $\delta_j$, i.e. $\tau_\gamma := 2\delta_i\delta_j/(\delta_i+\delta_j)$.
Note that the bilinear form $a+J$ is symmetric.
This variant of a DG-FEM approach is also called symmetric weighted interior penalty Galerkin (SWIPG) method \cite{Ern2008}.
\changedReview{For a sufficiently large $\eta>0$, the problem has a unique solution.}
Note that each integral can be replaced by a sum over element local contributions.

In the following, we will use $\PP_1(E)$ on tetrahedral and $\QQ_1(E)$ on hexahedral elements for the local polynomial spaces in $V_h^k$.
The polynomial space of each element $E_j\in\Tt_h$ is spanned by $N_b\in\NN$ local basis functions $\varphi_0^j,\dots,\varphi_{N_b-1}^j$ ($N_b=4$ for tetrahedrons, $N_b=8$ for hexahedrons).
\changedReview{On the reference element, these local basis functions can be given as $1, x, y, z$ for tetrahedrons and $1, x, y, z, xy, xz, yz,xyz$ for hexahedrons, respectively.}
By mapping each local basis function on each element to a unique global index, we obtain the global basis functions $\varphi_0,\dots,\varphi_{n-1}$, where $n=N_b\cdot N$.
\changedReview{Note that the support of each global basis function spans only a single element of the mesh.}

\subsection{An unfitted discontinuous Galerkin (UDG-FEM) method for solving the EEG forward problem}
\begin{figure}[!t]
  \centering
  \newcommand{\vA}{0.5}
\newcommand{\vBC}{1.2}
\newcommand{\vD}{1.4}
\definecolor{outsidecolor}{rgb}{0.44,0.72,0.63}
\definecolor{insidecolor}{rgb}{0.91,0.58,0.46}
\begin{tikzpicture}[scale=3]
  \fill[outsidecolor] (0,0) rectangle (1,1);
  \begin{scope}[domain={\vBC/(\vBC+\vD)}:1.0, samples=21]
    \fill[insidecolor] (1,1) -- plot (\x,{(\vA-(\vBC+\vA)*\x)/(\vBC+\vA-(\vA+2.0*\vBC+\vD)*\x)});
    \draw[thick] plot (\x,{(\vA-(\vBC+\vA)*\x)/(\vBC+\vA-(\vA+2.0*\vBC+\vD)*\x)});
  \end{scope}
  \draw[thick] (0,0) rectangle (1,1);
\end{tikzpicture}
\hspace{0.5cm}
\begin{tikzpicture}[scale=3]
  \begin{scope}[shift={(1.0,0)}]
    \fill[outsidecolor] (0,0) rectangle (1,1);
    \fill[insidecolor] (1,1) -- (1,{\vBC/(\vD+\vBC)}) -- ({\vBC/(\vD+\vBC)},1);
    \draw[thick] (1,{\vBC/(\vD+\vBC)}) -- ({\vBC/(\vD+\vBC)},1);
    \draw[thick, dashed] (0,0) -- (1,{\vBC/(\vD+\vBC)});
    \draw[thick, dashed] (0,0) -- ({\vBC/(\vD+\vBC)},1);
    \draw[thick] (0,0) rectangle (1,1);
  \end{scope}
\end{tikzpicture}
  \caption{Construction of the subtriangulation for a single bilinear level set function.
    In the left image, two domains are delimited by a bilinear level set function and in the right image, the resulting discrete domains are shown. The dashed lines delimit the elements of the subtriangulation which are only used for integration.}
  \label{fig:subtriangulation}
\end{figure}
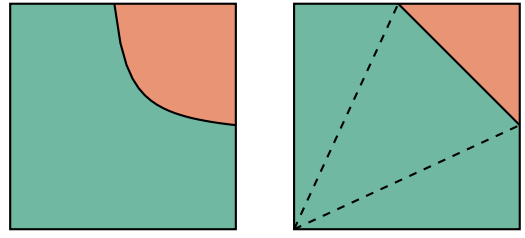

The former version of the DG-FEM approach uses a triangulation\changedReview{,} which resolves the geometry.
The unfitted discontinuous Galerkin (UDG-FEM) method takes a different approach and describes the geometry using level set functions.
We assume that the head domain is embedded in a larger domain $\hat\Omega\subset\RR^3$.
A level set function for a subdomain $\Omega\subset\hat\Omega$ is a scalar continuous function $\Phi:\hat\Omega\to\RR$ with the property
\begin{align}
  \Phi(x)
  \begin{cases}
    <0&\changedReview{\text{if }}x\in\Omega\changedReview{,}\\
    =0&\changedReview{\text{if }}x\in\partial\Omega\changedReview{,}\\
    >0&\changedReview{\text{if }}x\in\hat\Omega\setminus\overline{\Omega}\changedReview{.}
  \end{cases}
\end{align}
As an example, a level set function for the unit sphere can be defined as $\Phi(x) = \|x\|-1$.
On a structured grid $\Tt_h$, the level set function is approximated as a piecewise multilinear $\QQ_1$ function $\Phi_h$ by evaluating $\Phi$ at each grid node. 
In the following, we will call this structured mesh \emph{fundamental mesh}.
By employing multiple level set functions $\Phi_h^0,\dots,\Phi_h^{L-1}, \changedReview{L\in\NN}$, we can differentiate between multiple domains.
\changedReview{A level set can represent the boundary between two tissue compartments or delimit the domain of a single compartment (see e.g. \cite{Han2004}).}
Each such level set function separates $\hat\Omega$ into two parts with respect to its sign.
We will denote these parts by $\Omega_{i,-}$ and $\Omega_{i,+}$ for the negative and positive side of the level set function $i$.
\changedReview{From these parts, we can create domains $\Dd_0,\dots,\Dd_{D-1}, D\in\NN$, consisting of intersections of negative and positive sides of the level set functions, i.e. $\Dd_j\subset\{ \bigcap_{(i,p)\in I}\Omega_{i,p}: I\subset\lbrace 0,\dots,L-1\rbrace\times\lbrace -,+\rbrace \}$ for \changedReview{$j\in\lbrace 0,\dots,D-1\rbrace$}. Each domain can represent a tissue compartment.}
\changedReview{For a domain $\Dd_j$, we define its support $\Omega(\Dd_j)$ as the union of these intersections, i.e. $\Omega(\Dd_j):=\bigcup_{\tilde\Omega\in\Dd_j} \tilde\Omega$}. We require that the supports of all domains are pairwise disjoint, i.e., $\Omega(\Dd_i)\cap\Omega(\Dd_j) = \changedReview{\varnothing}$ holds for $i\not=j$.
For each element $E_i$, we set its set of intersecting domains to $\Dd(E_i):=\lbrace \Dd_j : E_i\cap\Omega(\Dd_j)\not=\changedReview{\varnothing}\rbrace$. 
From this domain information\changedReview{,} we generate a cut cell triangulation:
\begin{align}
  \label{eq:cutcelltriangulation}
  \overline{\Tt}_h=\bigcup_{E_i\in\Tt_h}\lbrace \underbrace{E_i\cap\Omega(\Dd_j)}_{=:E_i^j}: \Dd_j\in\Dd(E_i)\rbrace
\end{align}
The elements \changedReview{$E_i^j$} of $\overline{\Tt}_h$ are called \emph{cut cells}.
\changedReview{$E_i^j$ is thus defined as the part of element $E_i$ belonging to domain $\Dd_j$.}
In the following, we will require that the conductivity tensor $\sigma$ is constant on each cut cell.
On $\overline{\Tt}_h$\changedReview{,} we can again define a broken polynomial space and pose the DG problem \eqref{eq:dg}.
The local polynomial spaces are now defined on each cut cell.
They can be obtained by simply restricting the local basis function of the element to the cut cell, i.e. setting their value to zero outside of the cut cells domain.
\changedReview{The support of the global basis functions thus spans only a single cut cell.}
The integrals in \eqref{eq:dg} thus reduce to integrals of the local basis functions over cut cells.
Note that an element of the fundamental mesh can contain multiple sets of local basis functions, depending on the number of different domains in that element.
 An example can be seen in the left part of \figurename{} \ref{fig:subtriangulation}, where,  restricted to a single element of the fundamental mesh, two domains are separated by a bilinear level set function, thus forming two cut cells.
For each of these two cut cells, a set of local basis functions is introduced.

For the evaluation of integrals over a cut cell and its boundary, several methods have been proposed \cite{Papadopoulo2007}.
We use an extended marching cubes algorithm \cite{Engwer2016}.
The domain of a cut cell is approximated by a first order subtriangulation into simple elements.
As the discrete level set functions $\Phi_h^i$ are $\QQ_1$ functions, they are completely defined by their values at the mesh nodes.
The subtriangulation can be computed automatically solely based on these values.
The computation uses precomputed look up tables and produces topologically correct triangulations.
An example for such a subtriangulation in 2D can be seen in the right part of \figurename{} \ref{fig:subtriangulation} (the subtriangulation of the green subdomain is marked by the dashed lines).
For a detailed description of the extended marching cubes algorithm, we refer to \cite{Engwer2016}.
When an element is cut by multiple level set functions, we apply the marching cubes algorithm recursively on the elements of the subtriangulation.
Note that we introduce a slight error in the interface reconstruction by this recursive application.
However, this error decreases with decreasing $h$ and should be negligible \changedReview{when using linear basis functions.}
The integration over a cut cell $E$ can be replaced by integrations over the simple elements $E^j$ of the subtriangulation:
\begin{align}
  \int_{E} g(x) dx &= \sum_{j}\int_{E^j}g(x)dx\\
  &=\sum_{j}\int_{\hat E^j} g\left(\mu^j(\hat x)\right)\left|\operatorname{det}\left(J\mu^j(\hat x)\right)\right| d \hat x
\end{align}
\changedReview{with the affine map $\mu^j:\hat E^j\mapsto E^j$ mapping local coordinates of the reference element $\hat E^j$ to global coordinates of the element of the subtriangulation $E^j$.}
The local integrals over the reference elements of the subelements can be carried out using common quadrature rules.
Besides the subtriangulation of the volume, the marching cubes algorithm also produces a subtriangulation of the interface.
This can be used to compute the corresponding skeleton integrals.

When inserting the global basis functions into the bilinear form and the right hand side \changedReview{\eqref{eq:dg}}, we get a sparse linear equation system $My=f$, \changedReview{with $M\in\RR^{n\times n}$ and $y,f\in\RR^{n}$ where $n=N_b\cdot N$ denotes the global number of degrees of freedom.}
$M=(m_{ij})$ and $f=(f_{ij})$ are defined as
\begin{align}
  m_{ij}= a(\varphi_{i},\varphi_{j})+\frac{\eta}{h}J(\varphi_i,\varphi_j),\quad f_{i} = f(\varphi_i)
\end{align}
Since all basis functions have support in at most one element or cut cell, the entries of $M$ and $f$ can be blocked element or cut cell wise.
After solving this linear system for $y$, the potential $u$ is given as $u(x) = \sum_i y_i\varphi_i(x)$.

\section{Methods}
\label{sec:methods}

\subsection{Source model}
\label{subsec:sourcemodel}

For standard CG-FEM, several different source models have been proposed, such as the partial integration \cite{CHW:Wei2000,Vallaghe2010,Vorwerk2012}, the Saint-Venant \cite{Mun2012,CHW:Med2015}, the Whitney or Raviart Thomas \cite{Tanzer2005, Pursiainen2011} or the subtraction approach \cite{Schimpf2002,Wolters2007b}.
In principle, the presented method is applicable to any given source model.
Due to its simplicity and accuracy when compared to the subtraction approach 
in DG-FEM simulations \cite{Engwer2015,CHW:Vor2016}, 
we use here the partial integration approach to model a mathematical dipole.
When tested with a basis function $\varphi_i$, the following right-hand side results\changedReview{:}
\begin{align}
  f_{i} =  f(\varphi_i)=
  \begin{cases}
    -M\cdot\nabla\varphi_i\changedReview{(x_0)} &\text{if }x_0\in \operatorname{support}(\varphi_i)\\
    0&\text{else}
  \end{cases}
\end{align}
For the standard DG-FEM discretization, the support of the discrete source model is completely contained in a single element.
For the UDG-FEM approach, the same holds with respect to a single cut cell.
This is especially advantageous when solving the EEG forward problem for many dipoles using a fast transfer matrix approach (see below).
Note that for tetrahedral elements, where we use the $\PP_1$ basis, the gradient of the basis function is constant.
In this case, the source model does not depend on the local position within its supporting element.
\changedReview{In addition, if the dipole lies exactly on a cell boundary, we shift it slightly towards the interior of one of the neighboring elements.}

\subsection{Transfer matrix}
\label{subsec:transmat}

In order to reduce the computational load when solving the EEG forward problem for many sources, we use a fast transfer matrix approach \cite{CHW:Wei2000,CHW:Gen2004,Mun2012} \changedReview{, which is also related to the adjoint method \cite{Vallaghe2010b}}.
In most cases, we are not interested in the potential in the whole volume conductor, but only in the potential difference at a set of $N_{e}$ electrode positions $p_1,\dots,p_{N_e}\in\RR^3$ with respect to a reference electrode $p_0$.
We will denote the potential differences by $U\in\RR^{N_e}$.
These values can be obtained, by multiplying the solution vector $y\in\RR^n$ with a restriction matrix $R\in\RR^{N_e\times n}$: $U = Ry$.
The entries of the restriction matrix $R=(r_{k,i})$ are given as $r_{k,i} = \varphi_i(p_0)-\varphi_i(p_k)$.
Replacing $y$ by $M^{-1}f$ we get $U=RM^{-1}f=Tf$, with the transfer matrix $T=RM^{-1}\in\RR^{N_e\times n}$.
$T$ can be computed by solving $MT^t=R^t$ (considering the symmetry of $M$).
The latter can be carried out for each column of $T^t$ and $R^t$ separately.
For a given discrete source model, the potential differences at the electrode positions can now be computed by a simple matrix vector multiplication.

\subsection{Implementation}
\label{subsec:implementation}

The method is implemented using the Distributed and Unified Numeric Environment\footnote{http://www.dune-project.org} (DUNE) \cite{Bastian2008a, Bastian2008b, Blatt2007, Engwer2012}.
DUNE is a general purpose, open source C++ library for solving partial differential equations using mesh-based methods.
It is extensible by providing a modular structure and offering generic interfaces and separation between data structures and algorithms.
For representing tetrahedral and hexahedral conforming meshes, we use the DUNE-ALUGrid module \cite{Dedner2014}.
The discretization of the partial differential equation uses the DUNE-PDELab module \cite{Bastian2010}.

\subsection{Solver methods}
\label{subsec:solvers}

In order to solve the different linear equation systems, we use different solvers for DG-FEM and UDG-FEM.
For the DG-FEM approach we use a conjugate gradient method with an algebraic multigrid preconditioner (AMG) as presented in \cite{Bastian2012}.
Within this multigrid method, an SSOR preconditioner is used as a smoother and a direct sparse solver is employed as a coarse solver \cite{Davis2004}.
\changedReview{We used a V-cycle for the recursive scheme and applied two pre- and two post-smoothing steps on each level.}
For the UDG-FEM approach, a similar AMG approach should in principle be possible but has not yet been implemented so far.
For UDG-FEM, we use here a conjugate gradient method with a block incomplete LU decomposition as a preconditioner on the cut cell blocks \cite{Blatt2007}.
\changedReview{The iteration for all methods was stopped at a relative reduction of the residual $\|f-My\|_2$ of $10^{-8}$.}

\subsection{Multi-layer sphere model}\label{subsec:sphere}
To \changedReview{verify} and evaluate the new approach, we compared the numerical solution to a quasi-analytical solution in a four compartment sphere model \cite{Munck1993}. The radii of the four layers and the corresponding conductivity values are shown in Table \ref{tab:sphere_radii_and_cond}.
\begin{table}[tbp]
  \renewcommand{\arraystretch}{1.5}
  \centering
  \caption{Sphere radii, tissue labels and conductivity values from outer to inner compartment.}
  \label{tab:sphere_radii_and_cond}
  \begin{tabular}{c|c|c|c|c}
    \emph{Radius}&92 mm& 86 mm& 80 mm& 78 mm \\
    \emph{Tissue}& skin & skull & CSF & brain\\
    \emph{Conductivity}&0.43 S/m& 0.01 S/m& 1.79 S/m& 0.33 S/m \\
  \end{tabular}
\end{table}
For the conductivity values, we followed the recommendations of \cite{Dannhauer2011}.

\subsection{Meshing aspects}
For studying the convergence behavior of the UDG-FEM approach as the number of elements is increased,
we generated a sequence of meshes with different resolutions. We started with a fundamental mesh with \changedReview{$N_d=2^4$} elements in each dimension
and refined the mesh by doubling the number of elements in each dimension. 
Tab. \ref{tab:udgmeshes} shows the resulting number of cut cells and degrees of freedom (DOF; $N_b=8$ DOFs on each cut cell).
We denote these meshes by the number of DOFs as: UDG 39k, UDG 218k and UDG 1335k (see Figs. \ref{fig:convergence},  \ref{fig:dgtetraudg} and  \ref{fig:dghexaudg}).
\changedReview{The mesh widths $h$ of the three fundamental meshes are approximately 12.13 mm, 6.06 mm and 3.03 mm respectively.}
\begin{table}
  \renewcommand{\arraystretch}{1.5}
  \centering
  \caption{Number of cut cells and degrees of freedom for the different UDG-FEM discretizations. \changedReview{Note that the number of degrees of freedom is given as $\text{\emph{DOFs}}=8\cdot\text{\emph{total}}$.}}
  \label{tab:udgmeshes}
  \begin{tabular}{c||c|c|c|c|c||c}
    \changedReview{$N_d$}      & \emph{skin} & \emph{skull} & \emph{CSF} & \emph{brain} & \changedReview{\emph{total}}  & \emph{DOFs} \\
    $2^{4}$ & 1.328       & 1.120        & 872        & 1.520        & 4.840   & 38.720      \\
    $2^{5}$ & 6.872       & 5.776        & 4.064      & 10.552       & 27.264  & 218.112     \\
    $2^{6}$ & 37.864      & 32.800       & 18.184     & 78.000       & 166.848 & 1.334.784   
  \end{tabular}
\end{table}

For a comparison study of the UDG-FEM approach with the DG-FEM approach presented by \cite{Engwer2015}, we constructed for the DG-FEM approach 
a conforming mesh of the multi-layer sphere model with tetrahedral elements such that both methods used a similar number of DOFs.
The conforming mesh for DG-FEM was generated using a constrained Delaunay tetrahedralization, as implemented in the tetgen software \cite{Si2015,Vorwerk2014}. The resulting tetrahedral DG-FEM mesh (DG tet 1447k had 1.446.804 DOFs, which is close to the 
highest mesh resolution used for the UDG-FEM (UDG 1335k).

Furthermore, we created a regular hexahedral mesh of the multi-layer sphere model with 2 mm resolution, resulting 
in 3.056.904 DOFs (DG hex 3057k). This regular hexahedral mesh approximates the smooth surfaces of the 
spherical compartments only in a staircase like manner. The goal here is to compare DG-FEM on a
 higher resolution regular hexahedral mesh (DG hex 3057k) with UDG-FEM using much lower resolution 
(UDG 39k) to show the significant contribution of UDG-FEM to more accurately represent smooth tissue surfaces and thus better approximate realistic head geometries.

\begin{figure*}[t]
  \centering
  \subfloat[\changedReview{Conforming tetrahedral mesh}]{\includegraphics[width=0.25\linewidth]{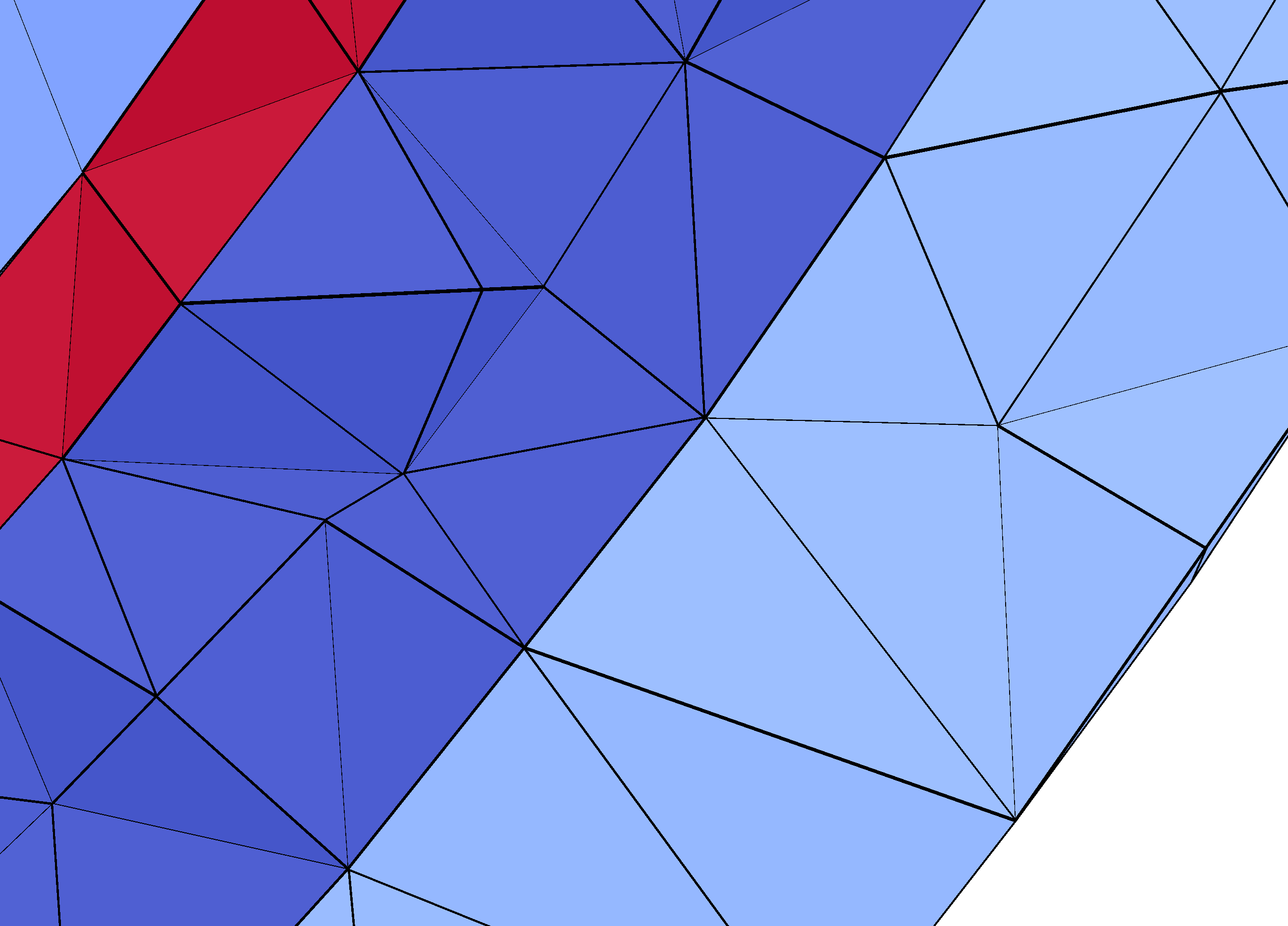}}\hspace{1cm}
  \subfloat[\changedReview{Conforming hexahedral mesh}]{\includegraphics[width=0.25\linewidth]{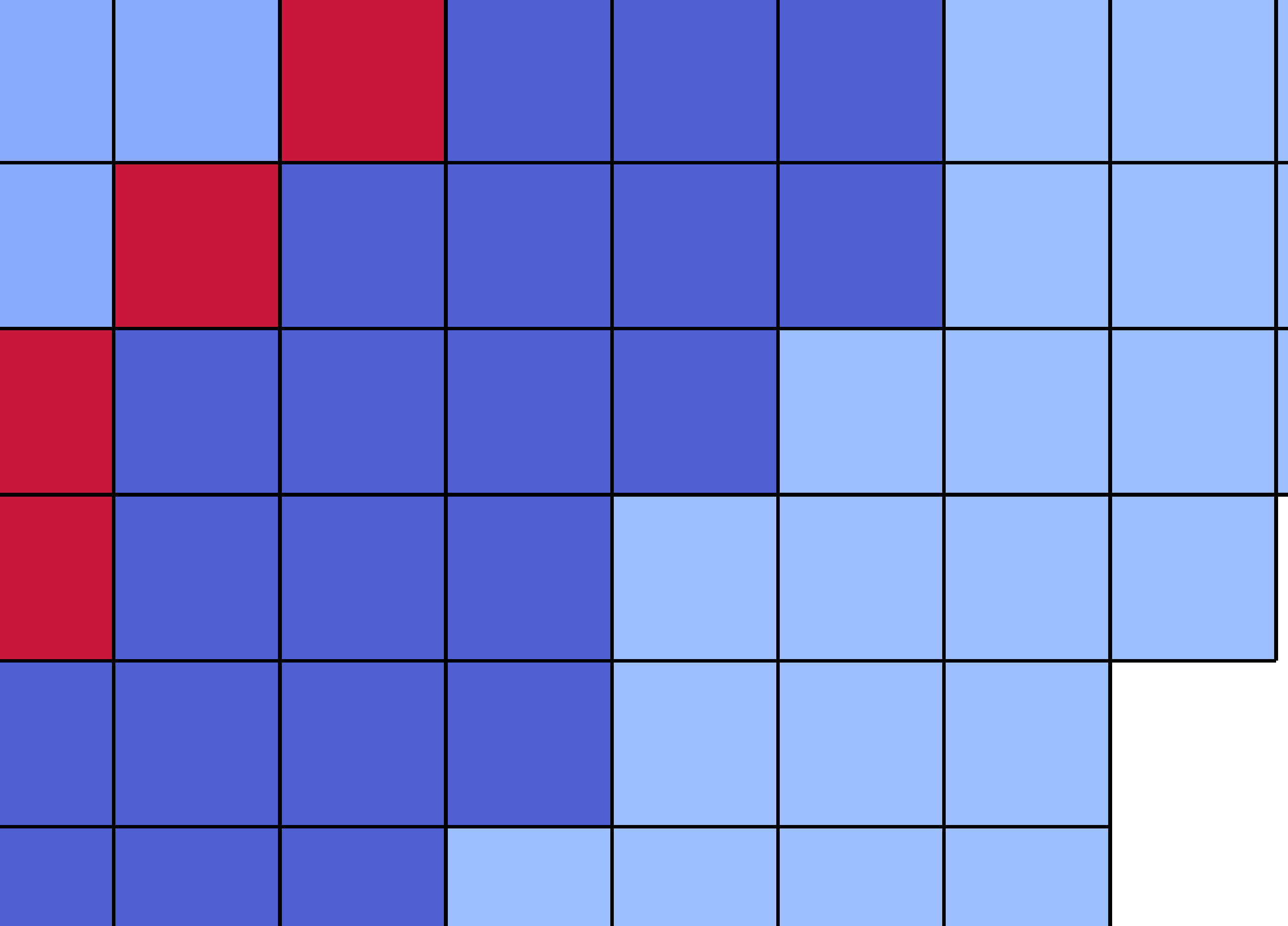}}\hspace{1cm}
  \subfloat[\changedReview{Cut cell mesh}]{\includegraphics[width=0.25\linewidth]{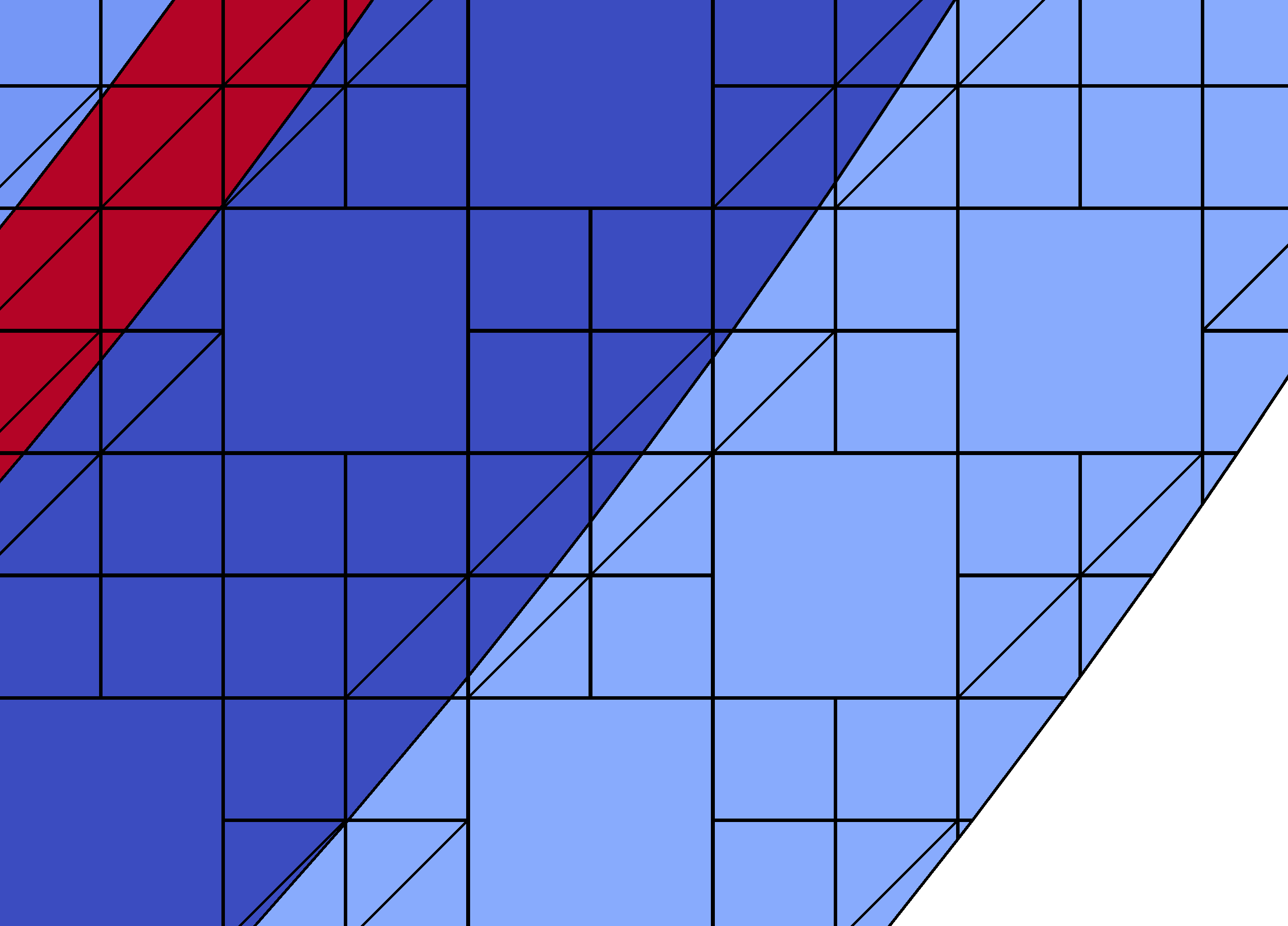}}
  \caption{\changedReview{Sections of the different meshes used in the multi-layer sphere verification. From left to right, the images show the \emph{DG tet 1447k}, \emph{DG hex 3057k} and \emph{UDG 1335k} models. The different colors represent the different conductivity values.}}
  \label{fig:sphere_meshes}
\end{figure*}
\changedReview{Fig. \ref{fig:sphere_meshes} shows sections of the conforming tetrahedral and hexahedral meshes as well as the finest cut cell mesh.}
For both methods and all meshes, we computed the transfer matrices and calculated the potentials at the electrodes for all sources 
as explained in the following.

\subsection{Sources}
We generated two sets of dipoles with unit strength, one with tangential and the other with radial orientation.
For each set, we generated 1000 dipoles at 10 different eccentricities in the inner compartment.
The location of the dipoles were computed randomly within the given eccentricity.
The orientation of the tangential dipoles were also chosen randomly in the tangential plane.
An eccentricity value of 0 denotes the center of the sphere and a value of 1 a location at the boundary of the inner compartment.
Because it is well known that numerical errors increase with increasing eccentricity (a reasoning for
this effect has been given in \cite{Wolters2007b}), the chosen eccentricities were scaled logarithmically 
with increasing eccentricity and range from 0.1666 to 0.9939.
The latter corresponds to a distance of only 0.48 mm to the inner sphere surface and thus
very high eccentricity (where thus also higher numerical errors have to be expected \cite{Wolters2007b}).
With regard to the application, distances between 1 mm (eccentricity of 0.9872) and 2.5 mm (0.9678) or 3 mm (0.9615)
seem to be the most important, because, depending on the location, the cortex has a thickness of about 2 to 6 mm \cite{CHW:Li2014}
and source locations should be chosen to be in the middle of the grey matter compartment. Among the 10 chosen eccentricities, the values between 0.82 mm (0.9895) and 2.45 mm (0.9686) come closest to this range, so that we will later focus on these values and on especially the middle value of this range, i.e., 1.42 mm (0.9818). 

\subsection{Error measures}
The analytical and numerical potential solutions for each dipole were evaluated at $200$ electrodes on the 
outermost surface of the  sphere model, denoted by $\Uana\in\RR^{200}$ and $\Unum\in\RR^{200}$, respectively \changedReview{\cite{Munck1993}}.
The error between both solutions was computed using two common measures, the \emph{relative difference measure} 
($\rdmp$) and the \emph{magnitude error} ($\magp$) \cite{Vorwerk2012}:
\begin{align}
  \rdmp(\Uana,\Unum)&=50\left\|\frac{\Uana}{\|\Uana\|_2}-\frac{\Unum}{\|\Unum\|_2}\right\|_2\\
  \magp(\Uana,\Unum)&=100\left( \frac{\|\Unum\|_2}{\|\Uana\|_2}-1 \right)
\end{align}
The measures are scaled such that $0\le \rdmp\le100$ and $-100\le \magp<\infty$ hold.
Note that the accuracy of a method is better, the closer both measures are to zero.
Since the quality of the simulation is known to depend on the local mesh geometry as well as on the 
intra-element position \changedReview{of a dipole}, $\rdmp$ and $\magp$ are analyzed statistically for the above described
sample of sources as suggested in \cite{Vorwerk2012}, including results for the different source 
eccentricities in separate box-plots. This statistical analysis includes maximum and minimum, indicated by upper and lower error bars, and thereby the total range
(TR). Furthermore, it includes the interval between upper and lower quartile, i.e., 
the interquartile range (IQR), also known as the spread, which is marked by a box with a 
black dash showing the median.

\subsection{Realistic head model}
\begin{figure}[b]
  \centering
  \includegraphics[width=0.5\linewidth]{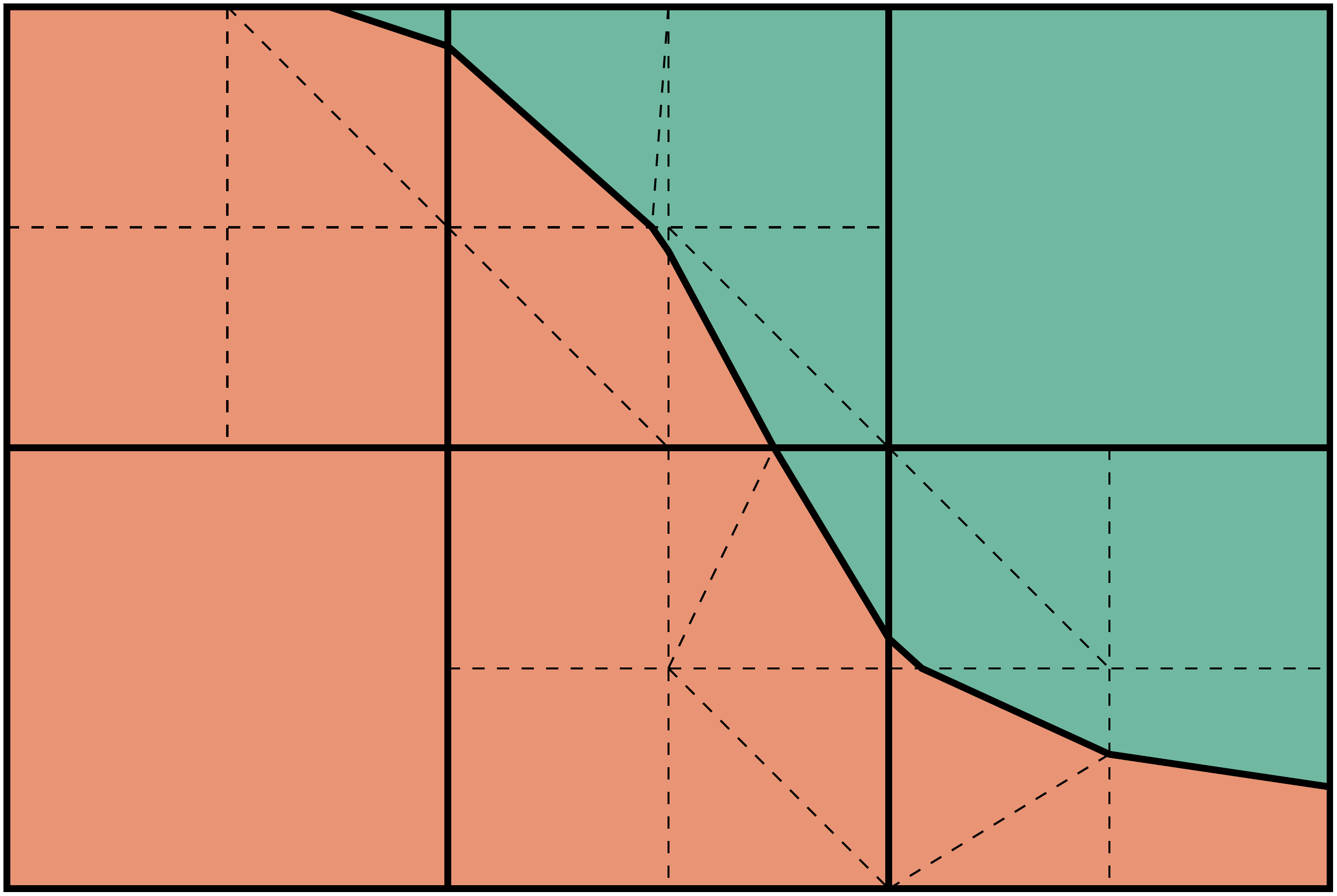}
  \caption{A section of the cut cell mesh for the realistic head model: the thick black lines show the cut cells on the fundamental mesh on which the basis functions are defined, while the dashed black lines depict the subtriangulation used for integration in the UDG-FEM approach.}
  \label{fig:arnocutcellmesh}
\end{figure}

To construct a realistic four compartment head volume conductor model\changedReview{,} we used the data of \cite{Vorwerk2014} with an adapted preprocessing pipeline:
T1-weighted (T1w-) and T2-weighted (T2w-) MRI scans of a healthy 25-year-old male subject were acquired in a 3 T MR scanner (MagnetomTrio, Siemens,
Munich, Germany) with a 32-channel head coil. For the T1w-MRI, an MP-RAGE pulse sequence (TR/TE/TI/FA = 2300 ms/3.03 ms/
1100 ms/8$\degree$, FOV = 256 $\times$ 256 $\times$ 192 mm, voxel size = 1 $\times$ 1 $\times$ 1mm) with fat suppression and GRAPPA parallel imaging (acceleration
factor = 2) was used. For the T2w image, an SPC pulse sequence (TR/TE = 2000 ms/307 ms, FOV = 255 $\times$ 255 $\times$ 176 mm, voxel size = 0.99
$\times$ 1.0 $\times$ 1.0 mm interpolated to 0.498 $\times$ 0.498 $\times$ 1.00 mm) was used. MR images were resampled to 1 mm isotropic resolution. 
The T2w-MRI was registered onto the T1w-MRI using a rigid registration approach and mutual information
as cost-function as implemented in the FSL-toolbox\footnote{http://fsl.fmrib.ox.ac.uk/fsl/fslwiki/FSL}. The skin and skull compartments 
were segmented by applying a gray-value based active contour approach \cite{Vese2002}. Subsequently, the segmentation
was manually corrected and, because of the importance of modeling skull holes for source analysis \cite{Oostenveld2002,Roche-Labarbe2008}, the foramen magnum and the two optic canals were correctly modeled as skull openings. 
The model was not cut off directly below the skull but realistically extended at the neck. 
CURRY7\footnote{http://www.neuroscan.com} was used to extract high-resolution surfaces of skin and skull compartments. 
A Taubin smoothing was applied to remove staircase-like effects \cite{Taubin1995}.
Cortex segmentation and surface reconstruction were performed using the FreeSurfer-toolbox\footnote{https://surfer.nmr.mgh.harvard.edu}.
From these surfaces, we computed signed distance functions on a 2 mm structured mesh, which were subsequently used as level set functions for defining the four different tissue compartments skin, skull, csf and brain.
The same conductivity values were used for these compartments as in the multi-layer sphere model.
The basis functions were defined on a mesh with a coarser resolution of 4 mm.
A section of the resulting mesh can be seen in \figurename{} \ref{fig:arnocutcellmesh}.
The resulting discretization had 106.031 cut cells and thus 848.248 DOFs and
skin, skull, csf and brain compartments consisted of 38.541, 25.587, 18.147 and 23.756 cut cells, respectively (head model REA 848k, see Fig. \ref{fig:realistic}).

\section{Results}
\label{sec:results}

\subsection{\changedReview{Verification} and evaluation in multi-layer sphere model}
\label{subsec:sphereresults}

\paragraph{Convergence of the UDG-FEM approach}
\begin{figure*}[!t]
  \subfloat[$\rdmp$ error for radial dipoles]{\includegraphics[width=0.49\linewidth]{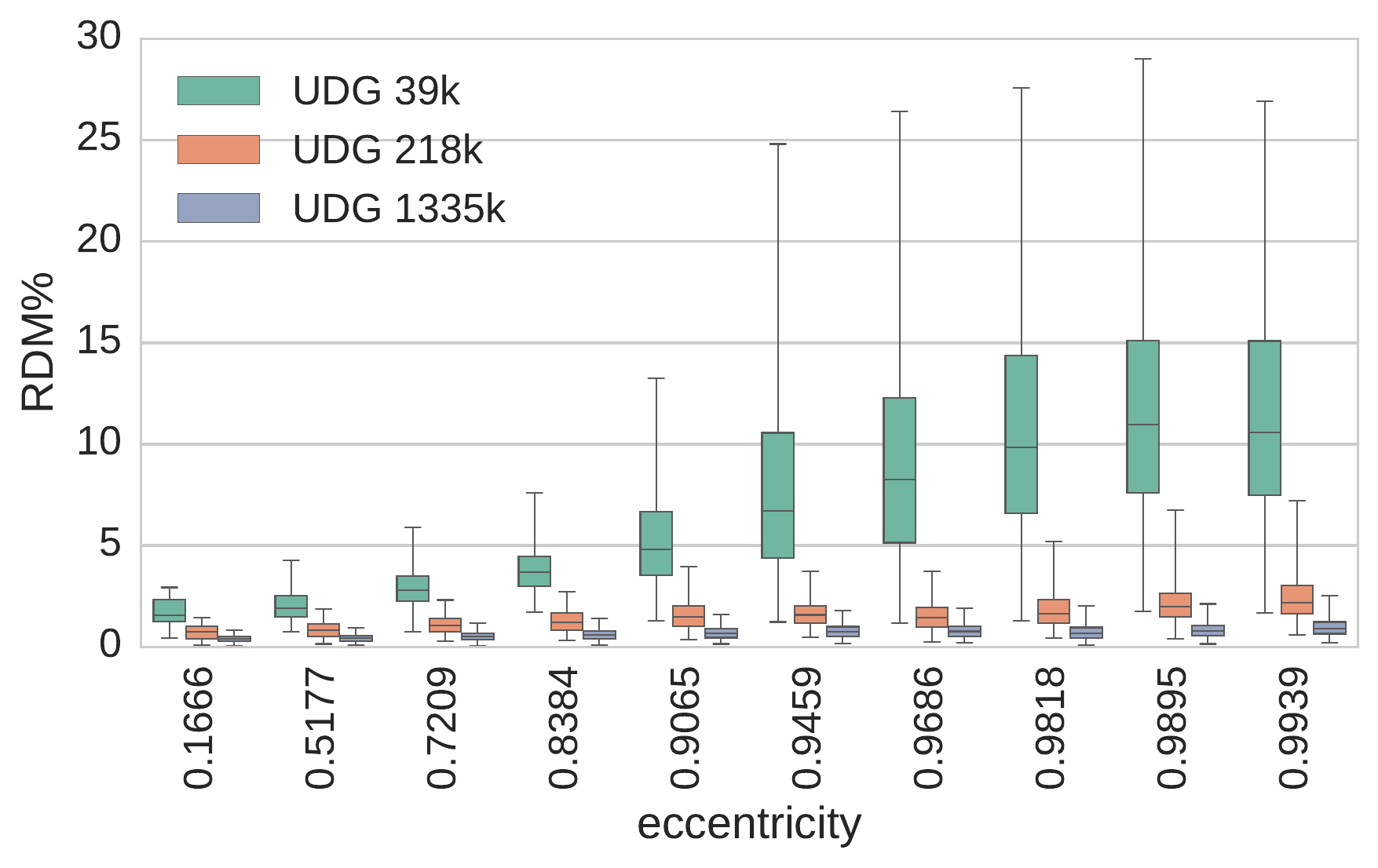}}
  \subfloat[$\rdmp$ error for tangential dipoles]{\includegraphics[width=0.49\linewidth]{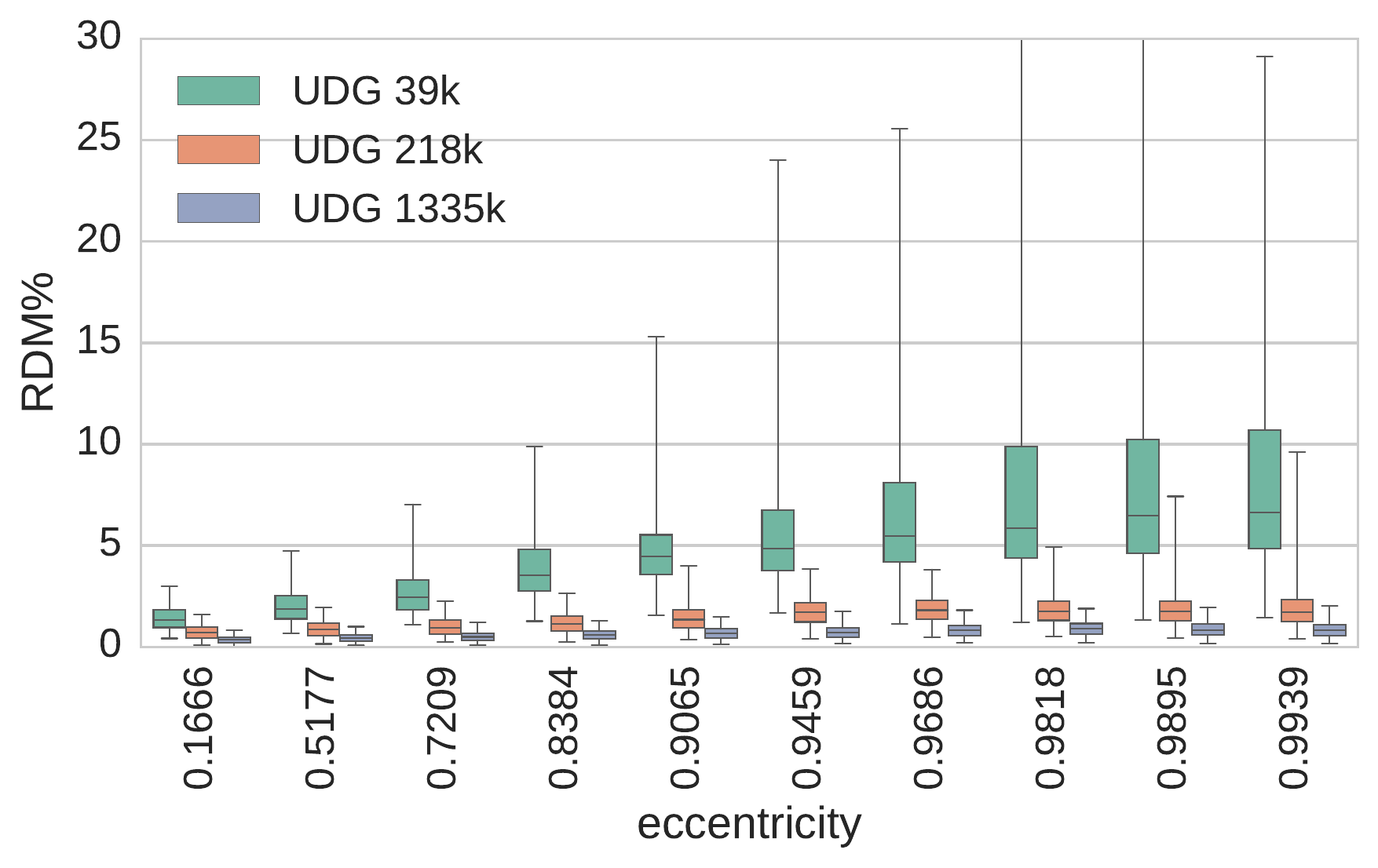}}\\
  \subfloat[$\magp$ error for radial dipoles]{\includegraphics[width=0.49\linewidth]{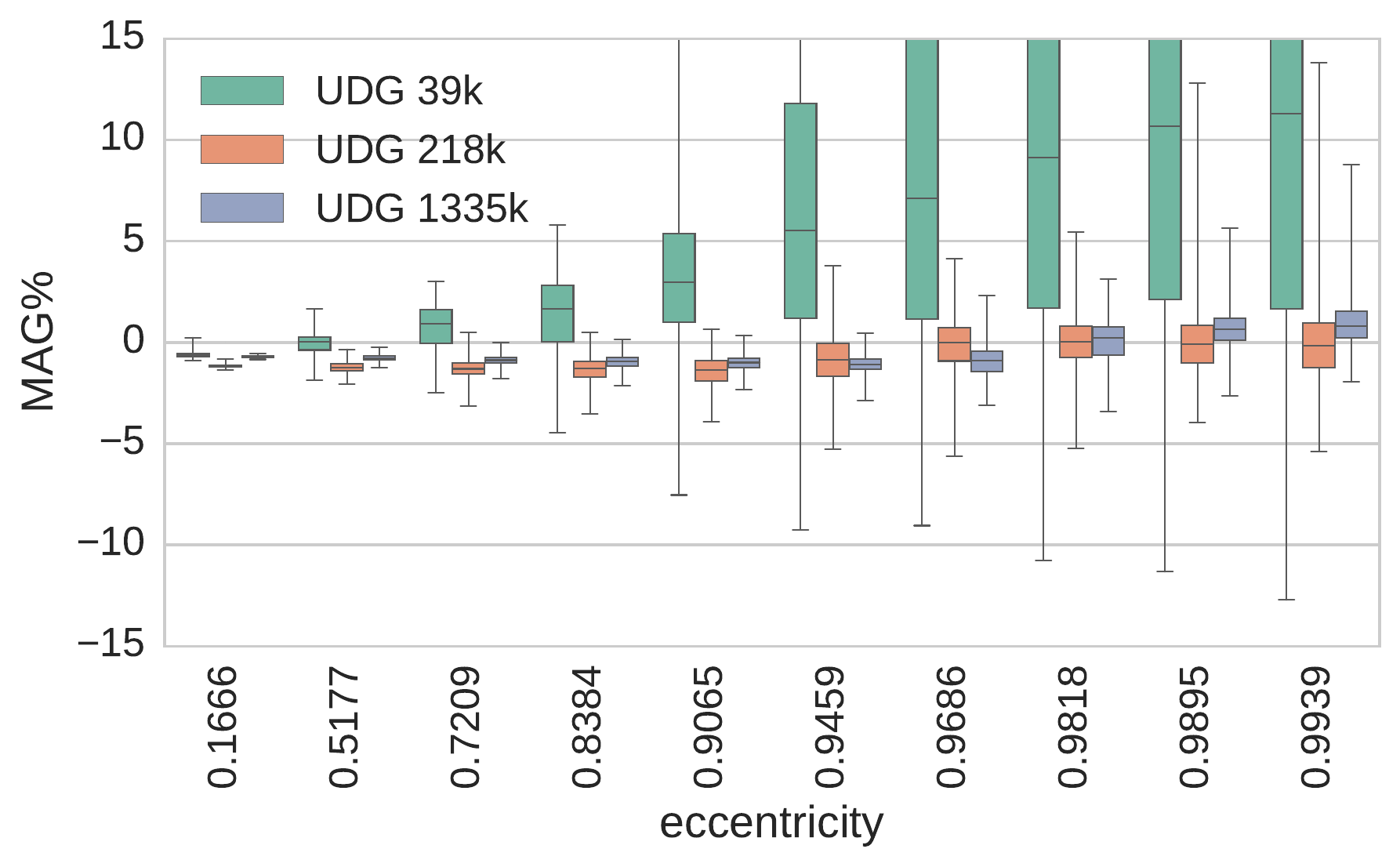}}
  \subfloat[$\magp$ error for tangential dipoles]{\includegraphics[width=0.49\linewidth]{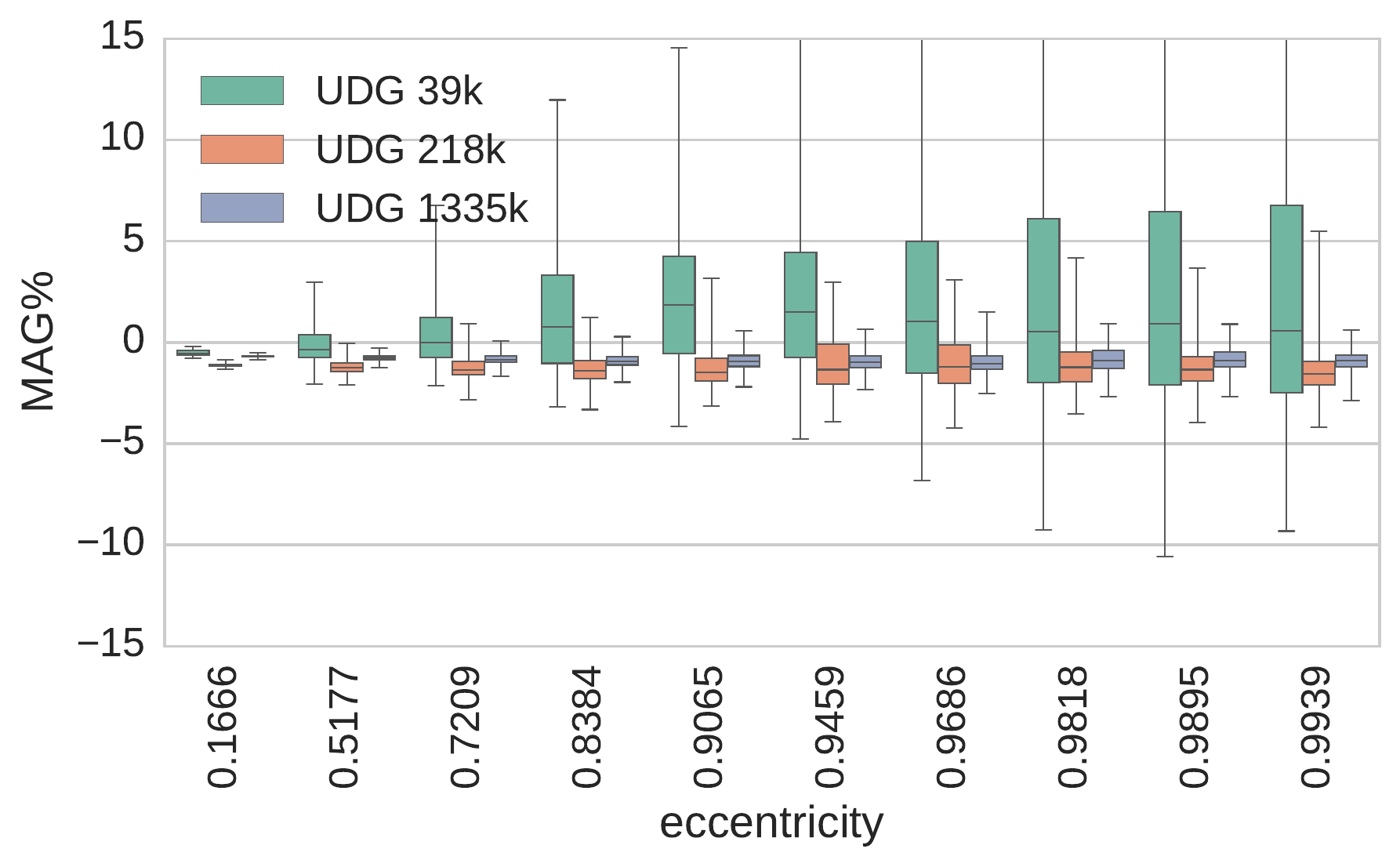}}
  \caption{ \changedReview{Verification} and convergence study for the UDG-FEM approach: $\rdmp$ (upper row) and $\magp$ (lower row) errors for radial (left column) and tangential (right column) sources for the three different resolutions 39k (green), 218k (red) and 1335k (blue).  Note that the x-axis is logarithmically scaled.}
  \label{fig:convergence}
\end{figure*}
In a first study, we \changedReview{verified} the new UDG-FEM approach and examined its convergence behavior for increasing mesh resolution.

The $\rdmp$ and $\magp$ statistical errors can be seen in \figurename{} \ref{fig:convergence} (note: x-axes are logarithmically scaled).
With increasing source eccentricity\changedReview{,} we observe an overall increase in the $\rdmp$ error and an increasing total range (TR) and interquartile range (IQR) of the $\magp$ error for all meshes.

With regard to the $\rdmp$, the finest mesh achieves a maximal error of 2.5 \% for radial sources and 2.0 \% for tangential sources.
Both of these values are observed at the maximal eccentricity of 0.9939 (i.e., only 0.48 mm from
the inner sphere surface). The median value over all eccentricities is 0.6 \% for radial and for tangential sources.
At the distance of 1.42 mm (eccentricity value 0.9818), from coarse to fine mesh, the IQR decreases from 7.7 \% over 1.1 \% down to 0.5 \% for radial sources and from 5.5 \% over 0.9 \% down to 0.5 \% for tangential sources. The TR behaves similarly.

For the $\magp$ error on the finest mesh, we observe a maximal absolute value over all eccentricities of 8.8 \% for radial and 3.4 \% for tangential sources.
For radial and tangential sources, the median value over all eccentricities is 0.8 \%.
When examining the $\magp$ at the distance of 1.42 mm (0.9818), from coarse to fine mesh, the 
IQR decreases from 16.5 \% over 1.0 \% down to 0.8 \% for radial sources and from 4.9 \% over 1.3 \% down to 0.8 \% for tangential sources.
The TR behaves again similarly.

\paragraph{Comparison of UDG-FEM to DG-FEM on conforming mesh with tetrahedral elements}
\begin{figure*}[!t]
  \subfloat[$\rdmp$ error for radial dipoles]{\includegraphics[width=0.49\linewidth]{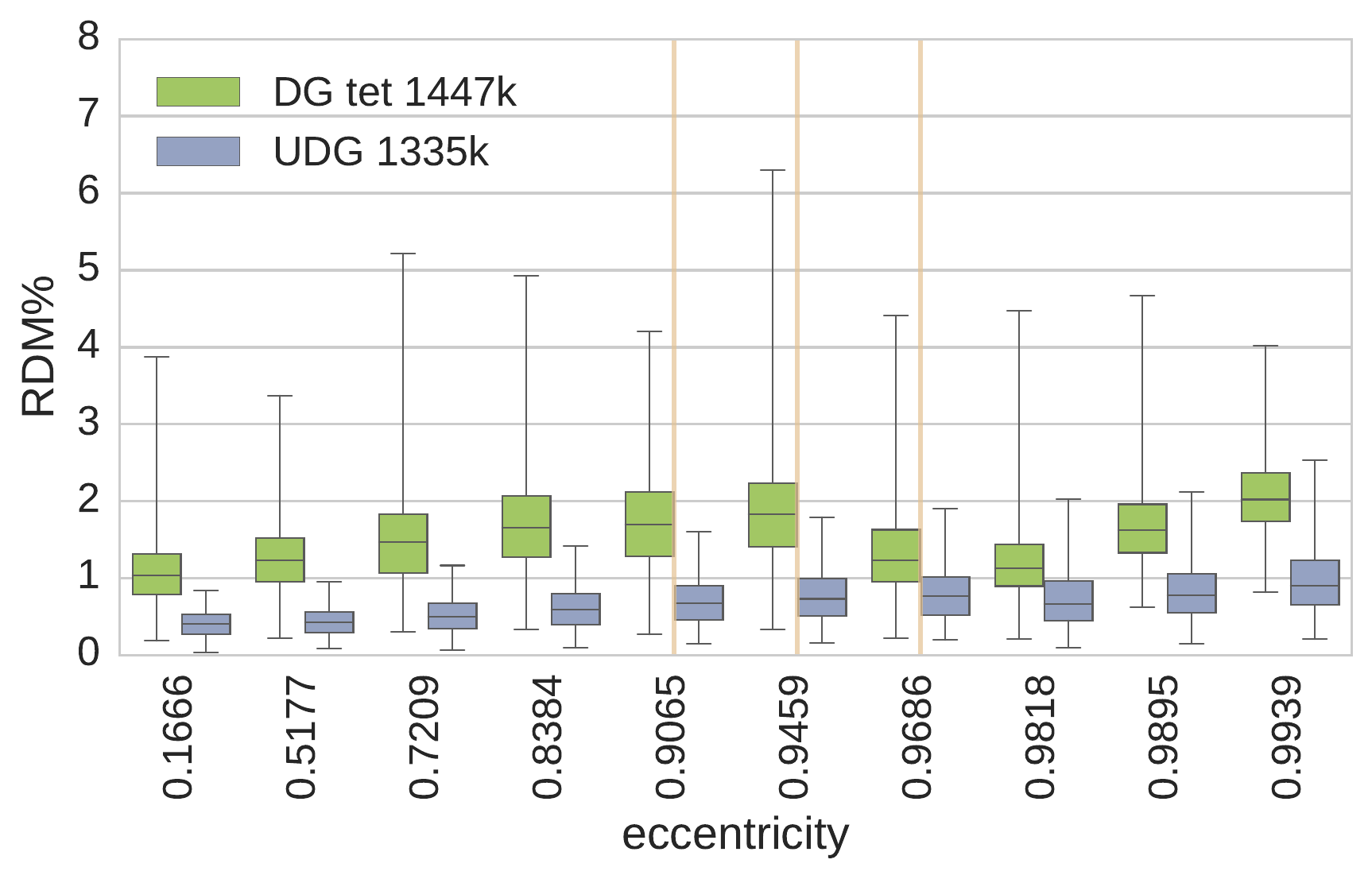}}
  \subfloat[$\rdmp$ error for tangential dipoles]{\includegraphics[width=0.49\linewidth]{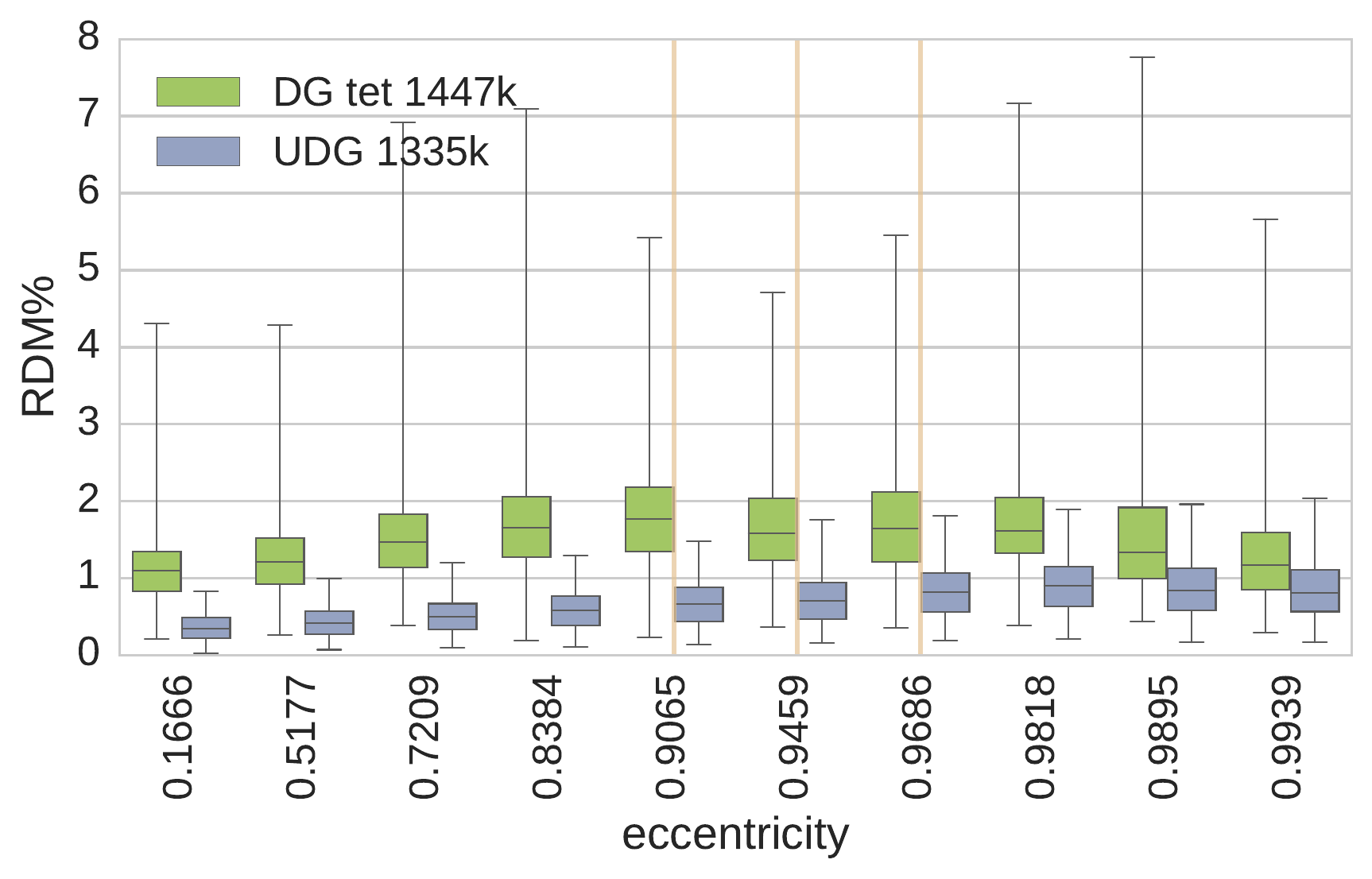}}\\
  \subfloat[$\magp$ error for radial dipoles]{\includegraphics[width=0.49\linewidth]{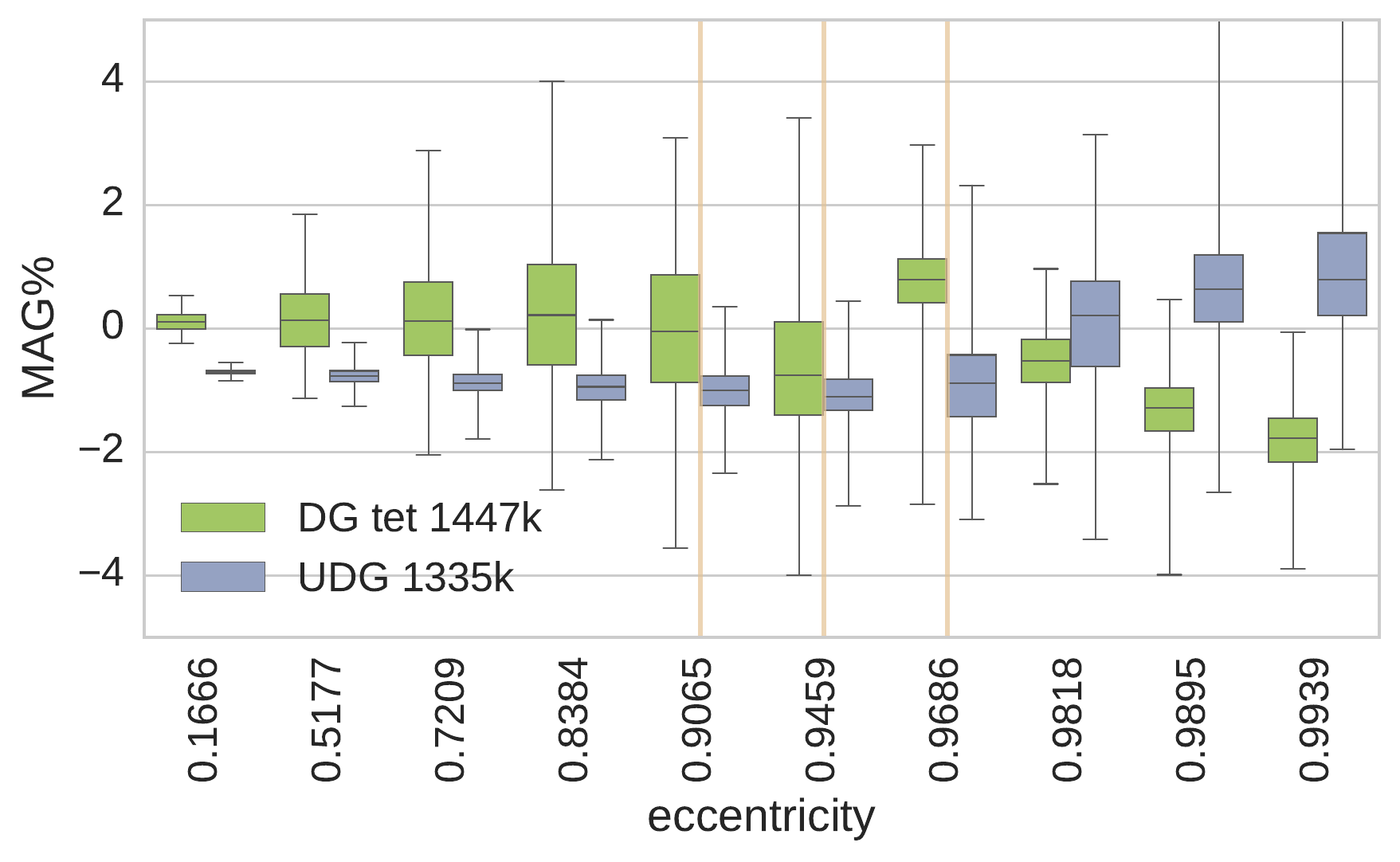}}
  \subfloat[$\magp$ error for tangential dipoles]{\includegraphics[width=0.49\linewidth]{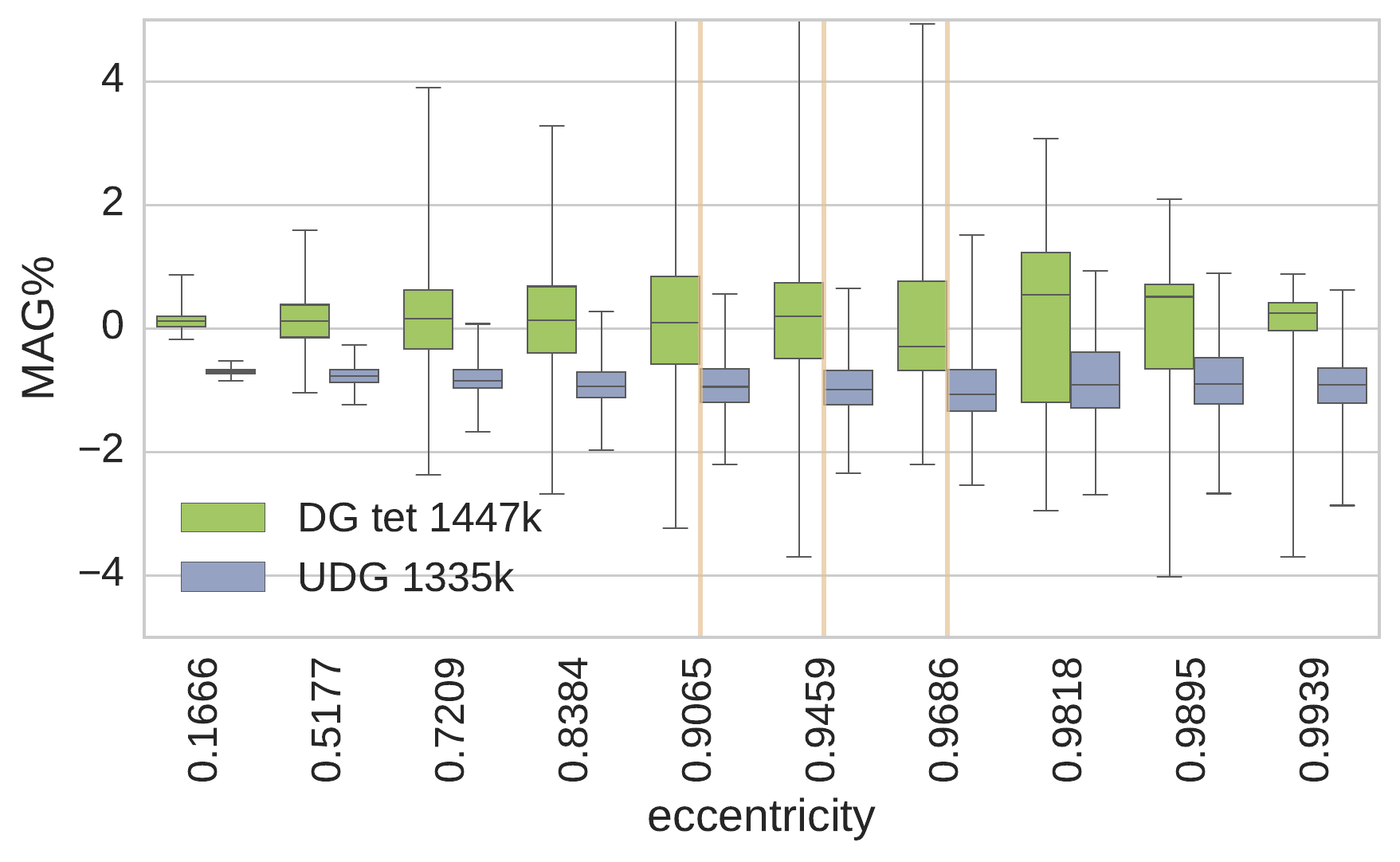}}
  \caption{Comparison between the UDG-FEM approach using model 1,335k (blue) and the DG-FEM approach on a conforming mesh with tetrahedral elements using model 1,447k (green): $\rdmp$ (upper row) and $\magp$ (lower row) errors for radial (left column) and tangential (right column) sources. Note that the x-axis is logarithmically scaled. With vertical \changedReview{yellow} lines, the maximal, mean and minimal (from left to right) element diameter is indicated of those elements in the inner compartment touching the compartment boundary. This diameter is given relative to the radius of the inner sphere. Note that this information is only relevant for the DG-FEM approach.}
  \label{fig:dgtetraudg}
\end{figure*}
In \figurename{} \ref{fig:dgtetraudg}, we compare the UDG-FEM approach using model UDG 1335k (blue) with the DG-FEM approach on the conforming mesh with tetrahedral elements using model DG tet 1447k (green). We show the $\rdmp$ (upper row) and $\magp$ (lower row) errors for radial (left column) and tangential (right column) sources (note: x-axes are again logarithmically scaled).
\changedReview{The source model for the DG-FEM approach with tetrahedral elements does not depend on the local 
position of the dipole within its supporting element, as the right hand side only involves the gradient of the basis functions, which is constant for linear functions.}
We therefore indicate by three vertical \changedReview{yellow} lines the maximal (6.70 mm at 0.9141), mean (4.77 mm at 0.9389) and minimal (3.27 mm at 0.9581) element diameter of those elements in the inner compartment which touch the brain/csf compartment boundary. 
\changedReview{Note that this information is only relevant for the DG-FEM approach with tetrahedral elements.}
As no dipoles were located at exactly those values, we indicated instead their closest eccentricities.

Since the UDG-FEM results were already discussed above, we focus now on the DG tet 1447k results of \figurename{} \ref{fig:dgtetraudg}.
We observe an overall increasing $\rdmp$ error up to the aforementioned left vertical line.
A similar observation can be made for the TR and IQR of the $\magp$ error. The maximal $\rdmp$ error over all eccentricities is 6.3 \% 
for radial dipoles and 7.8 \% for tangential dipoles. For both source orientations, the mean value over all eccentricities is 1.5 \%.
For radial sources, the maximal absolute value of the $\magp$ error is 4.0 \% and for tangential sources it is 5.3 \%.

Most importantly, with regard to $\rdmp$, over all eccentricities and for both source orientations, with an average median value of 0.7 \%, a TR of 1.5 \% and an IQR of 0.4 \%,  UDG 1335k performs better than DG tet 1447k,
which only achieves an average median value of 1.5 \%, a TR of 5.8 \% and an IQR of 0.8 \%. 
For the $\magp$ the situation is less clear with,  for  UDG 1335k, an average median value of -0.7 \%, a TR of 4.4 \% and an IQR of 0.8 \% and, for DG tet 1447k, an average median value of -0.2 \%, a TR of 6.0 \% and an IQR of 1.4 \%.

At the distance of 1.42 mm (0.9818), with regard to $\rdmp$, we find for radial sources for UDG-FEM a median of 0.7 \%, an IQR of 0.5 \%  and a TR of 1.9 \% and a median of 1.1 \%, an IQR of 0.5 \%  and a TR of 4.3 \% for the DG-FEM. For tangential sources, a median of 0.9 \%, an IQR of 0.5 \%  and a TR of 1.7 \% can be noted for UDG-FEM and a median of 1.6 \%, an IQR of 0.7 \%  and a TR of 6.8 \% for DG-FEM. 
With regard to $\magp$, we find for radial sources a median of 0.2 \%, an IQR of 1.4 \%  and a TR of 6.5 \% for UDG-FEM and a median of -0.5 \%, an IQR of 0.7 \%  and a TR of 3.5 \% for the DG-FEM. For tangential sources we find a median of -0.9 \%, an IQR of 0.9 \%  and a TR of 3.6 \%  for UDG-FEM and a median of 0.5 \%, an IQR of 2.4 \%  and a TR of 6.0 \% for DG-FEM. 

In summary, in the sphere model \changedReview{verification} study the UDG-FEM overall outperforms the DG-FEM.
Note in addition, that in the case of a realistic head, the model generation for the tetrahedral DG-FEM is far more sophisticated and, because touching brain and skull surfaces are not allowed, the volume conductor representation might be even less appropriate than it is for UDG-FEM. 

\paragraph{Comparison of low-resolution UDG-FEM and high-resolution DG-FEM on a structured mesh with hexahedral elements}
\begin{figure*}[!t]
  \subfloat[$\rdmp$ error for radial dipoles]{\includegraphics[width=0.49\linewidth]{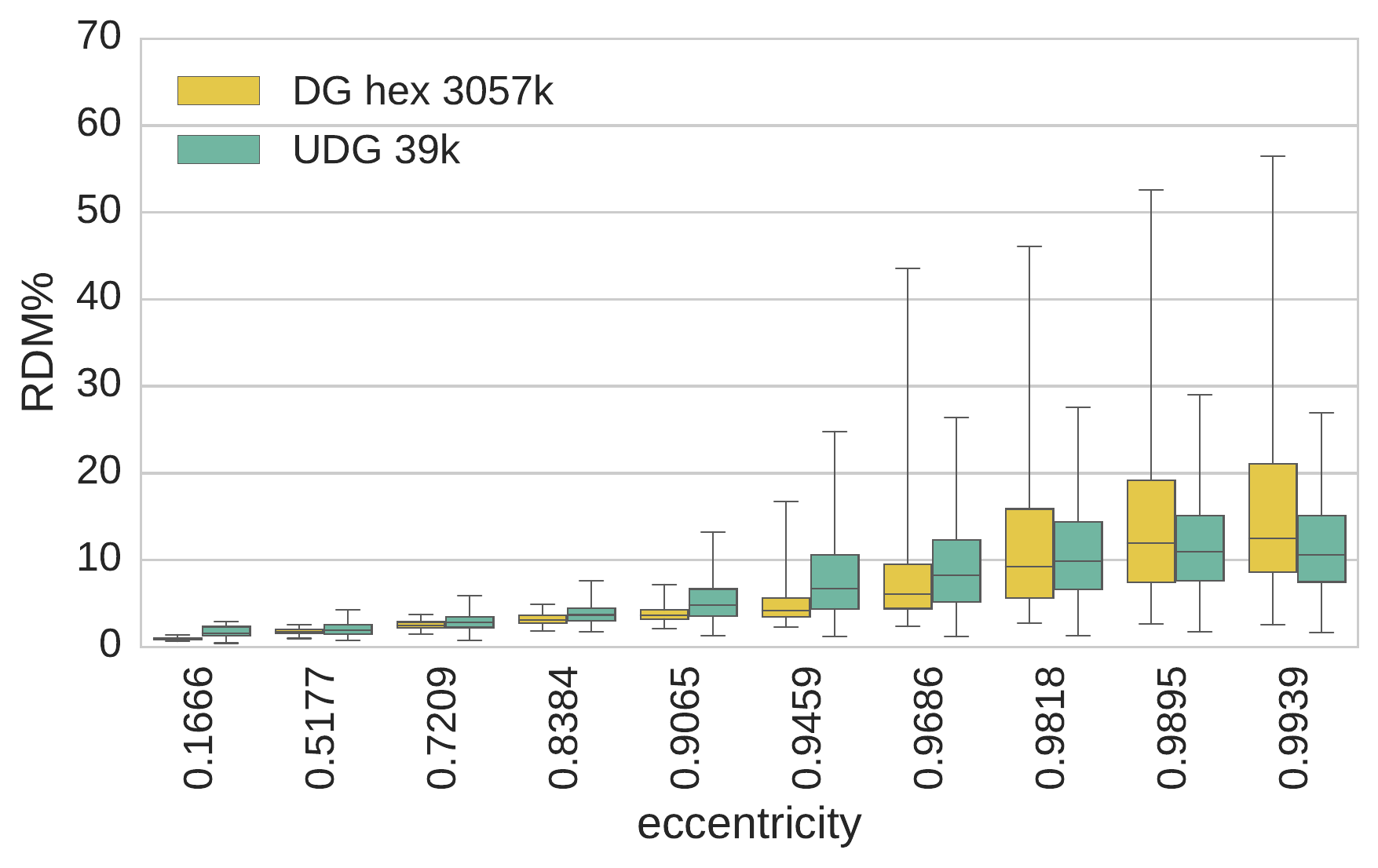}}
  \subfloat[$\rdmp$ error for tangential dipoles]{\includegraphics[width=0.49\linewidth]{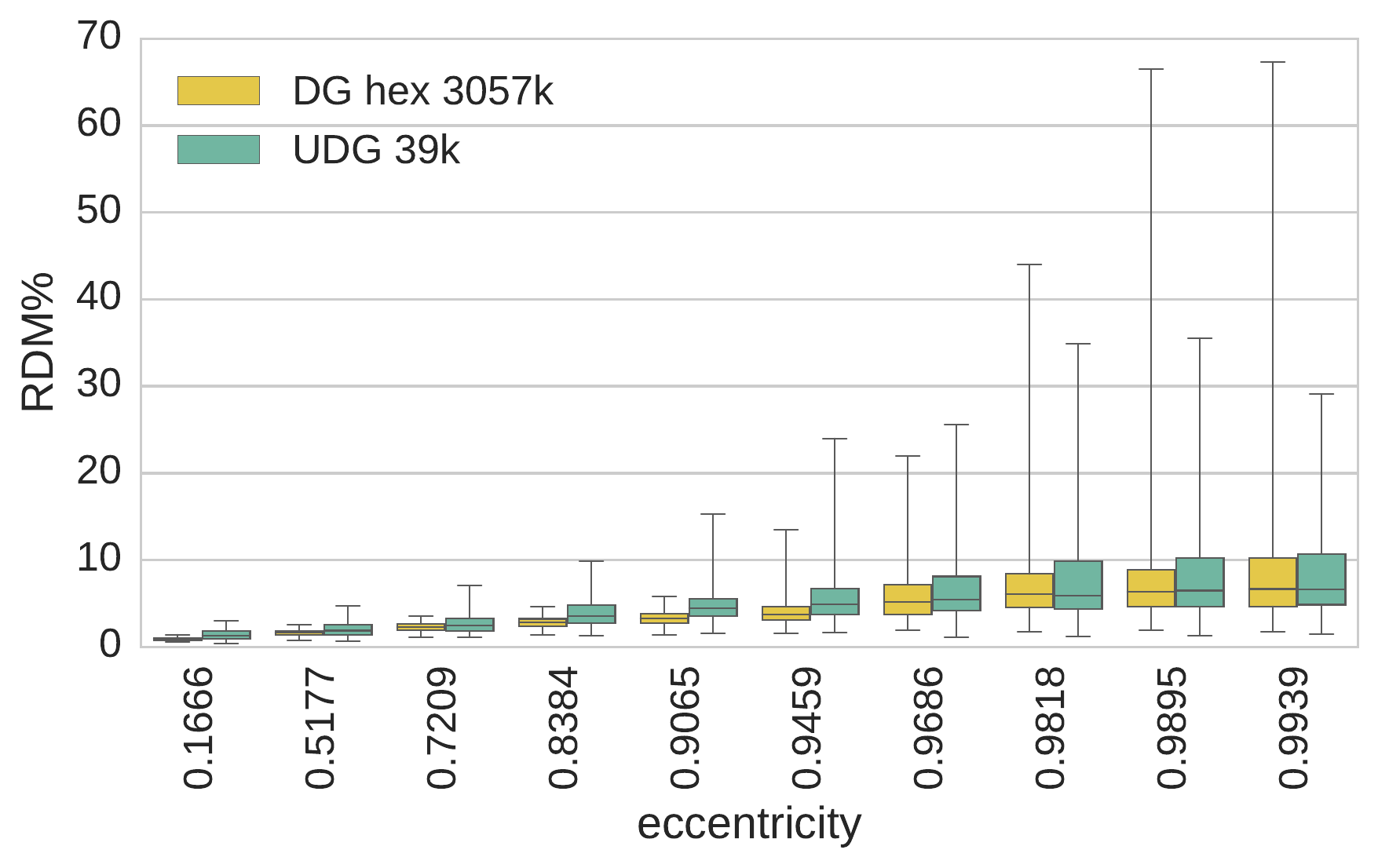}}\\
  \subfloat[$\magp$ error for radial dipoles]{\includegraphics[width=0.49\linewidth]{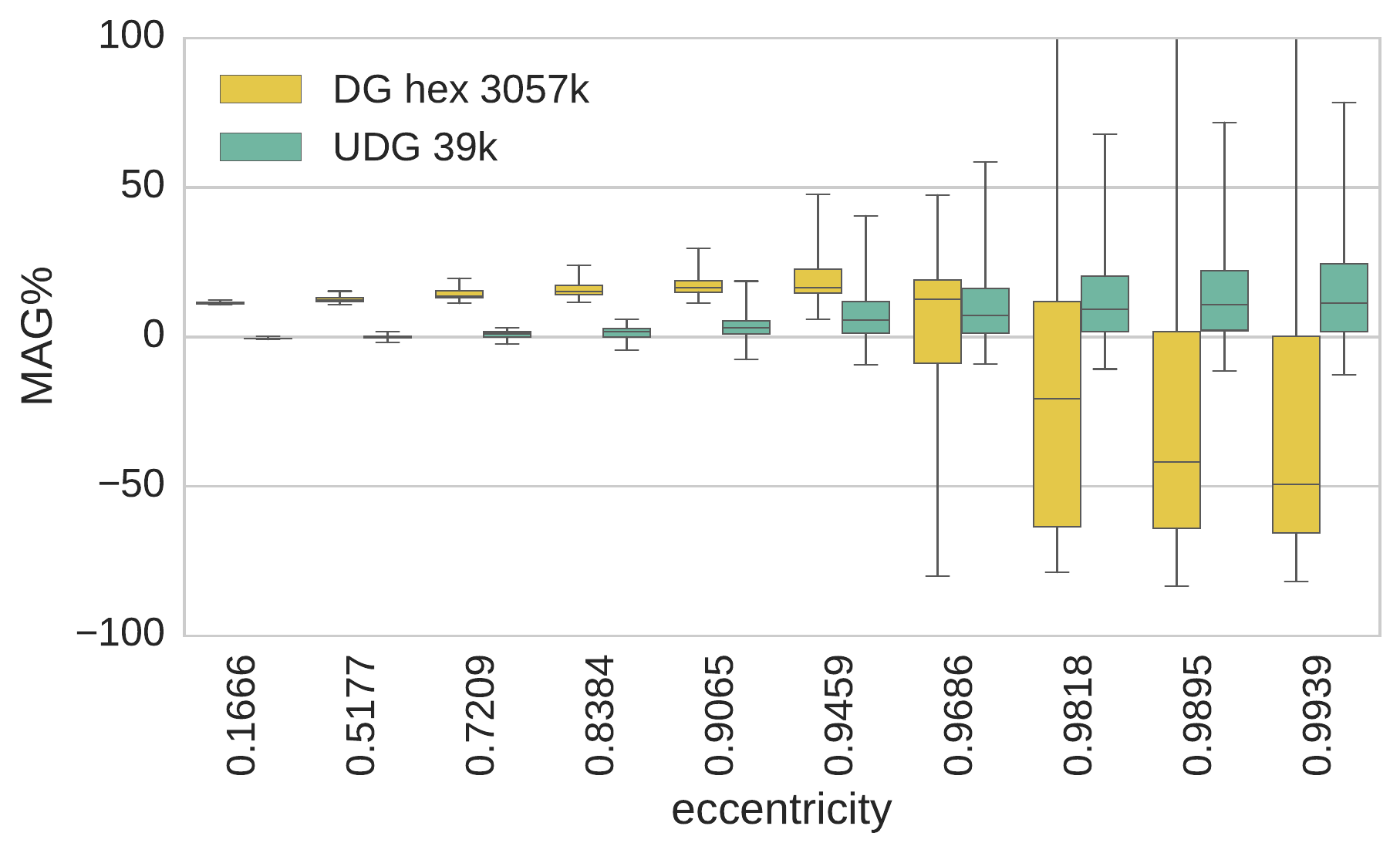}}
  \subfloat[$\magp$ error for tangential dipoles]{\includegraphics[width=0.49\linewidth]{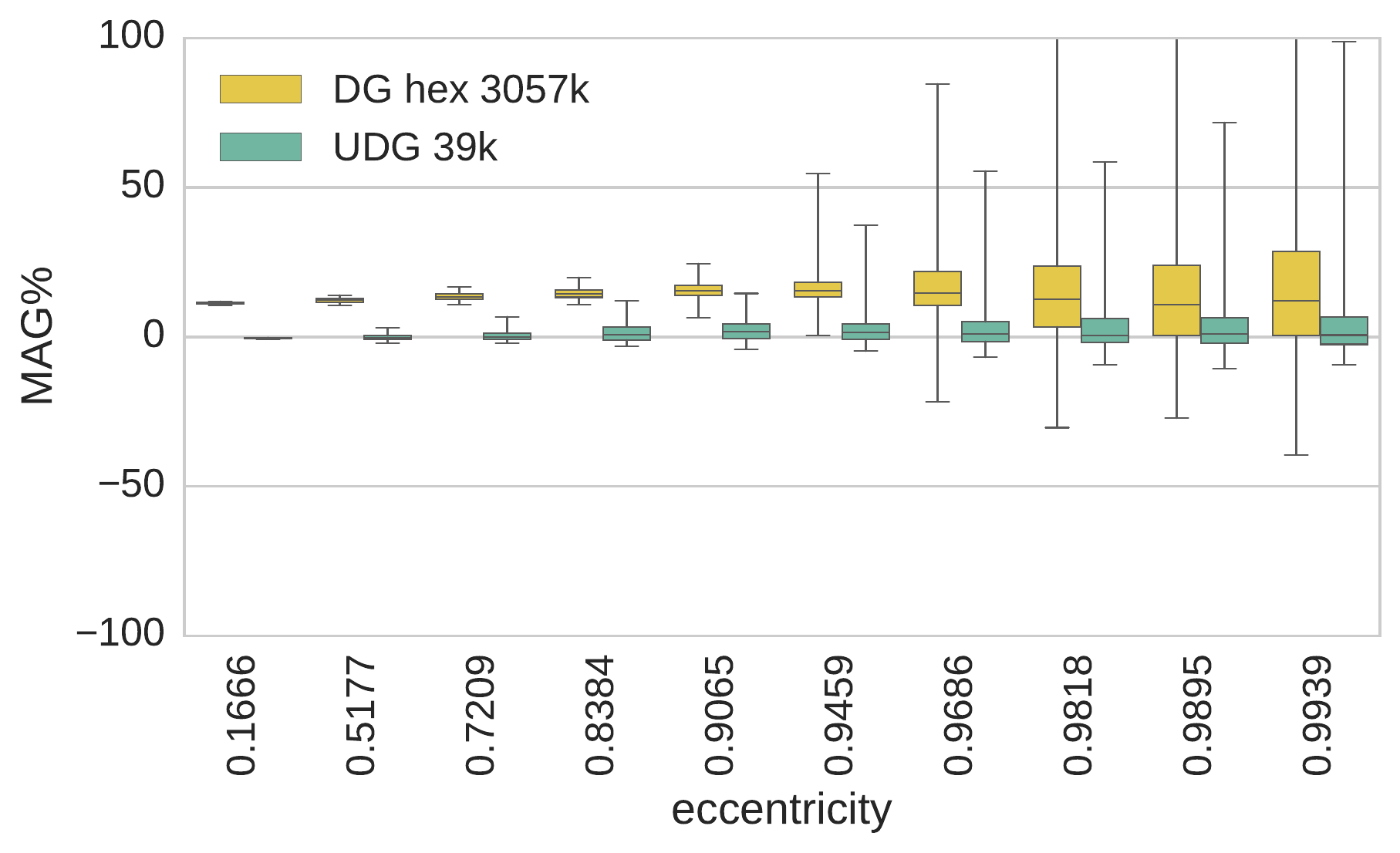}}
  \caption{Comparison between the UDG-FEM approach with 38.720 DOF (green) and the DG-FEM approach on a structured mesh with hexahedral elements and 3.056.904 DOF (yellow): $\rdmp$ (upper row) and $\magp$ (lower row) errors for radial (left column) and tangential (right column) sources. Note that the x-axis is logarithmically scaled.}
  \label{fig:dghexaudg}
\end{figure*}
In \figurename{} \ref{fig:dghexaudg}, we compare the UDG-FEM approach (green) with the DG-FEM approach on a structured mesh with 2 mm hexahedral elements (yellow) and show the $\rdmp$ (upper row) and $\magp$ (lower row) errors for radial (left column) and tangential (right column) sources. Note that the x-axes are again logarithmically scaled.
For the UDG-FEM approach\changedReview{,} we use the results from the coarsest mesh (39k DOFs) in the first test, while the DG-FEM approach uses about 79 times the number of DOFs (3057k).

With regard to $\rdmp$, even if for source eccentricities below 2.45 mm (0.9686) DG hex 3057k has slightly lower median, IQR and TR errors, for the more important higher eccentricities, both methods perform similarly.
With regard to $\magp$, the UDG-FEM performs even significantly better than the much higher resolution DG-FEM: For radial dipoles and source eccentricities below 2.45 mm (0.9686), the mean of all median values is 1.7 \% for UDG 39k, while it is 14.2 \% for DG  hex 3057k and for tangential sources, we find 0.5 \% and 13.7 \%, respectively. For radial sources at 1.42 mm distance to the csf (0.9818), the $\magp$ median error of DG hex 3057k is at -20.6 \%, while it is only at 9.1 \% in case of UDG 39k. Similarly, the IQR of DG hex 3057k of 75.3 \% and TR of 304.6 \% is reduced to 18.6 \% and 78.6 \% in case of UDG 39k, respectively.  For tangential sources, the median of 12.5 \% of DG hex 3057k is reduced to only 0.5 \% for UDG 39k, the IQR of 20.2 \% down to 8.1 \% and the TR of 142.8 \% down to 67.8 \% in case of UDG 39k.

In summary, the UDG-FEM with only 39k DOFs achieves already better or at least comparable results than the 2 mm regular hexahedral DG-FEM approach with about 79 times the number of DOFs (3057k). This test-case shows the high importance of a good geometry-approximation as performed by the UDG-FEM. With regard to application, higher resolution is needed to reduce the numerical errors especially for highly eccentric sources. The UDG approach thus has to be combined with a sufficient resolution, as also shown in the first study (\figurename{} \ref{fig:convergence}).

\subsection{Forward simulation in the realistic head model}
\begin{figure*}[!t]
  \centering
  \includegraphics[width=0.22\linewidth]{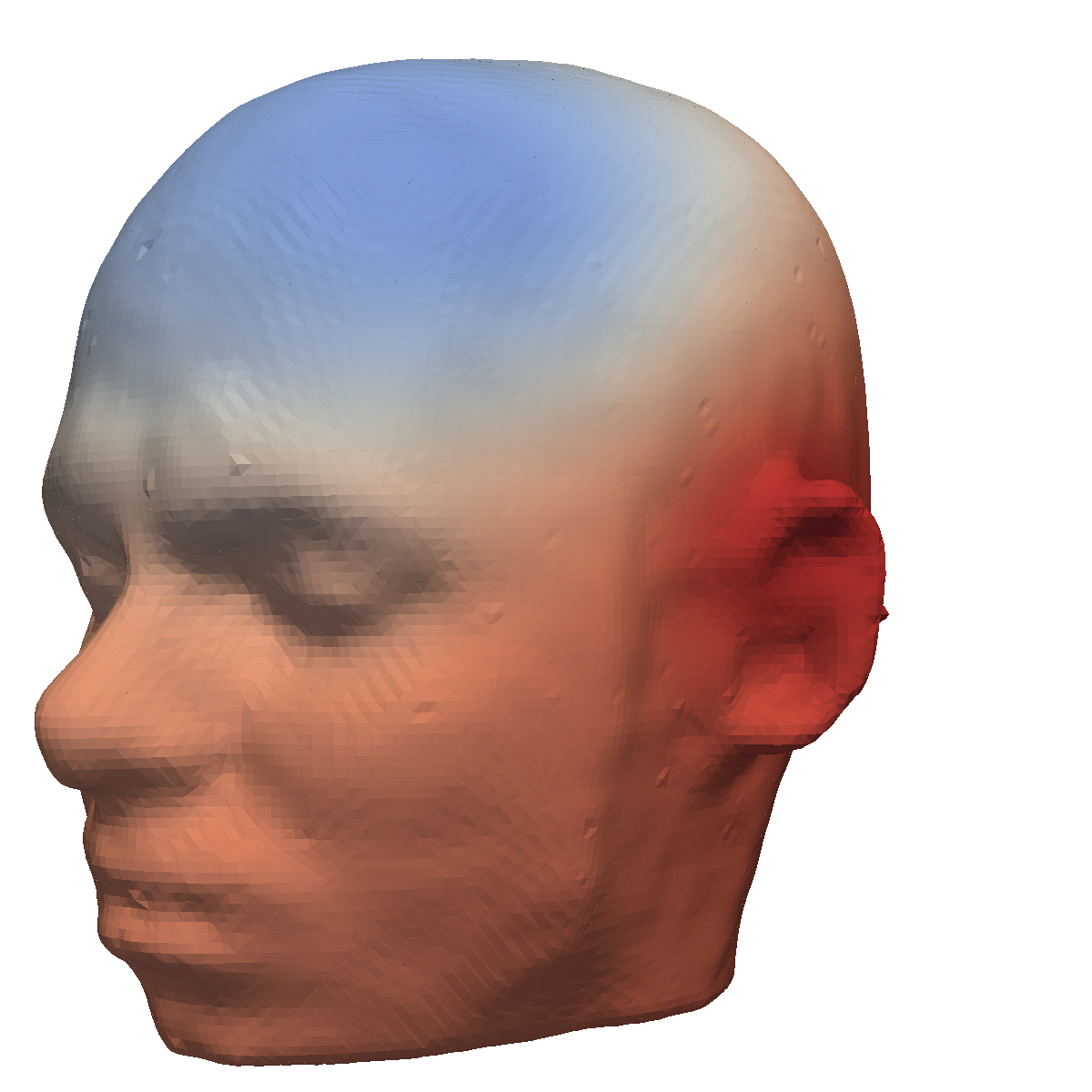}
  \includegraphics[width=0.22\linewidth]{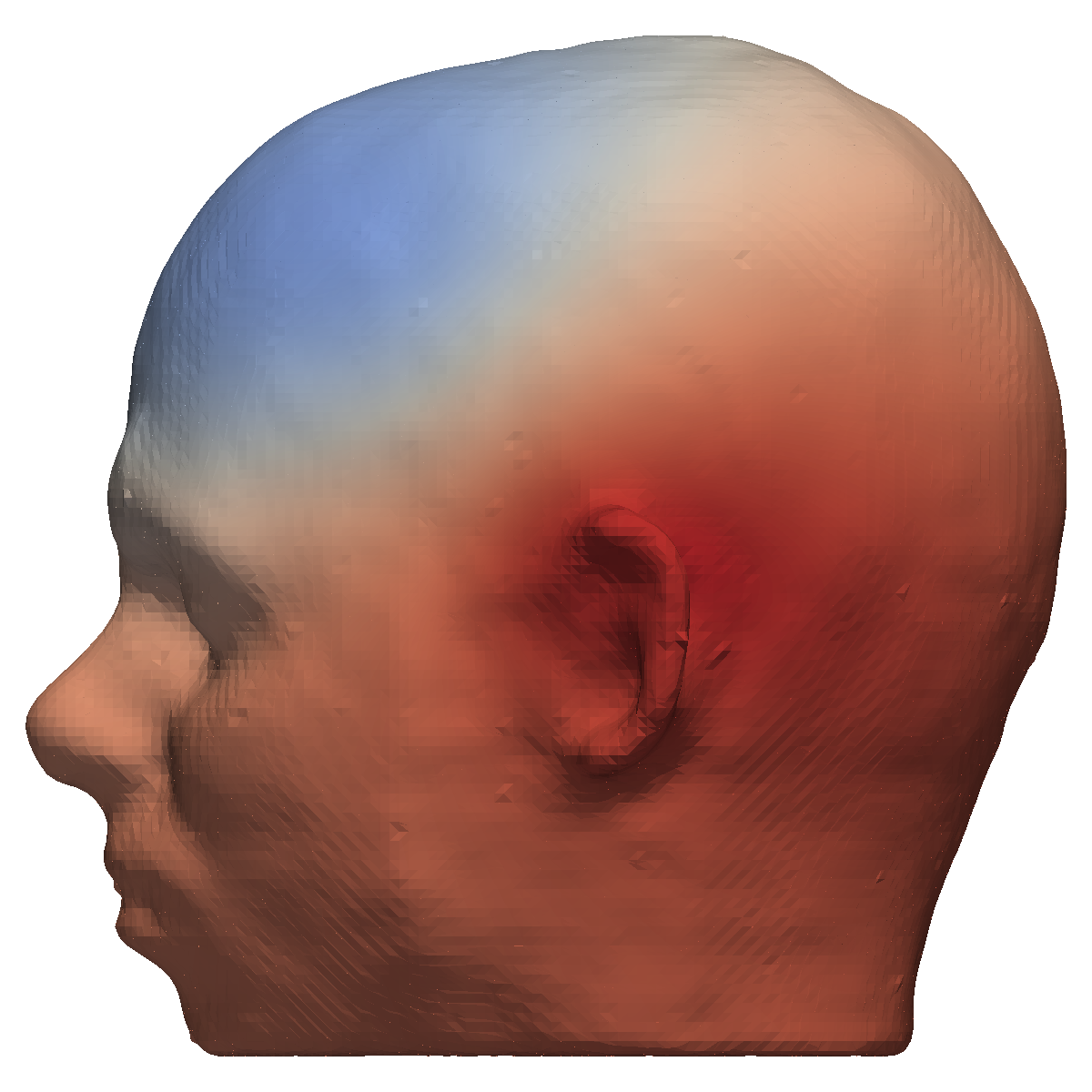}
  \includegraphics[width=0.22\linewidth]{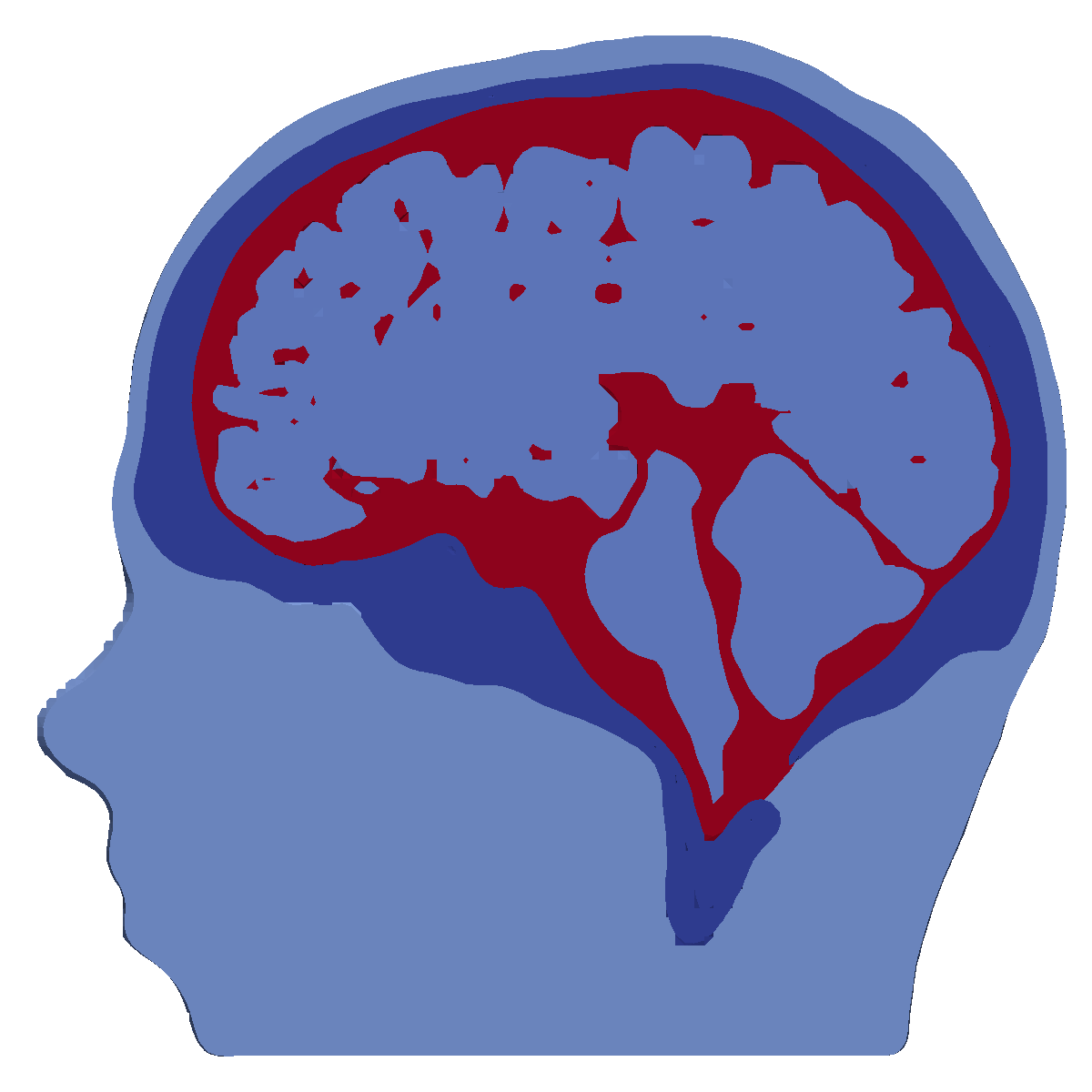}
  \includegraphics[width=0.22\linewidth]{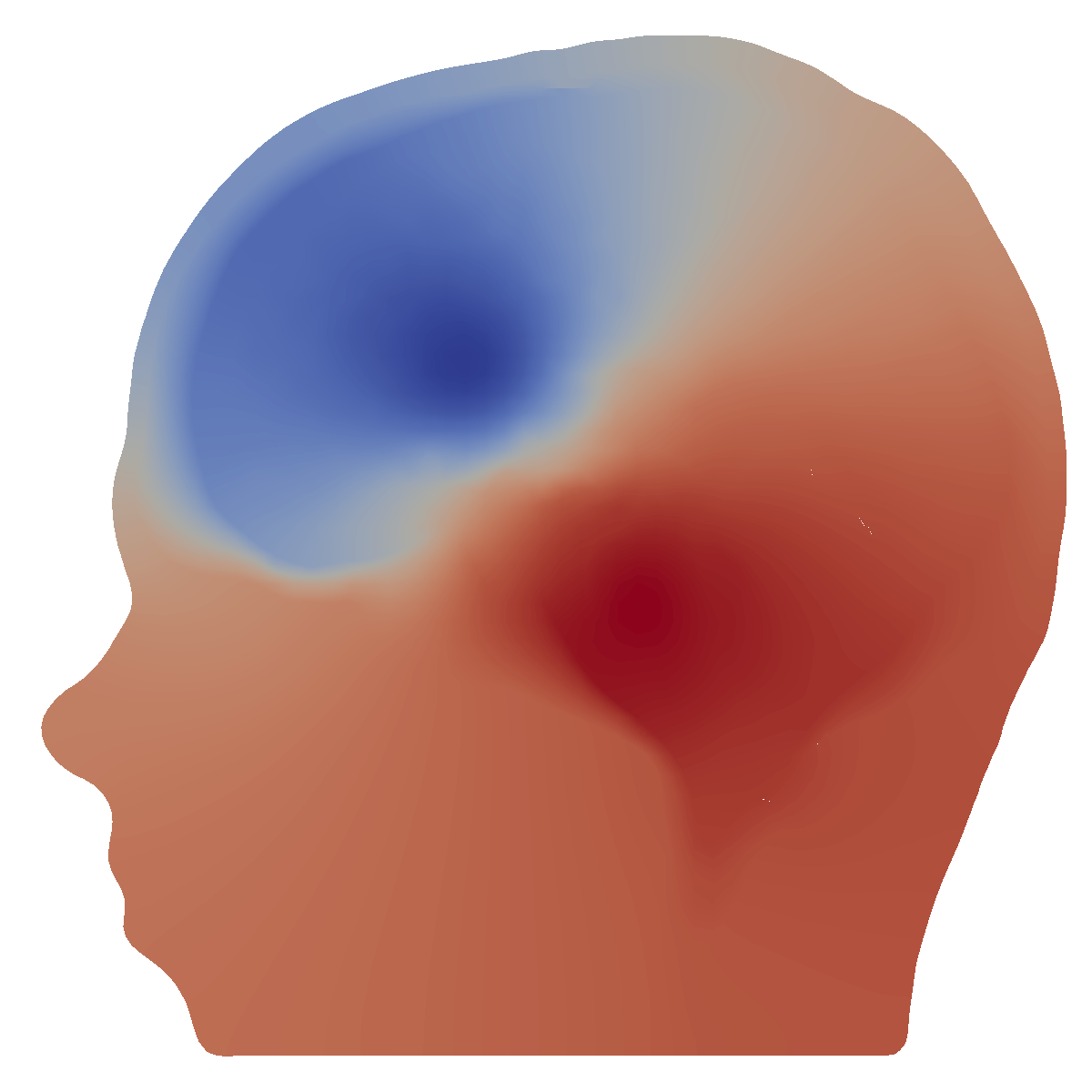}
  \caption{Exemplary EEG forward solution for an auditory source computed by UDG-FEM in the realistic four compartment head model REA 848k. The two images on the left show the potential at the scalp surface and the two images on the right the four tissue compartments and the potential on a sagittal slice through the head model.}
  \label{fig:realistic}
\end{figure*}
In the last study, we computed an example solution of the EEG forward problem with the UDG-FEM approach for a current dipole in the realistic four compartment head model REA 848k. The result is shown in \figurename{} \ref{fig:realistic}.
The current dipole is located in the auditory cortex and oriented normally to the grey matter surface.
The two images on the left show the potential distribution on the scalp surface and the two images on the right the different tissue compartments and the potential distribution on a sagittal slice through the head model.
A smooth representation of the different tissue surfaces, obtained from the level set functions, can be observed.
The forward simulation leads to a negativity in frontal areas and a positivity in the lower left neck area which is in line with practical findings for auditory evoked potentials \cite{Haemaelaeinen1993}.
\changedReview{The sequential runtime for the solution of this example simulation was approximately 2 minutes with a preceding 4 minutes of setup time on a single core of a conventional laptop.}

\section{Discussion}
\label{sec:discussion}

In this paper\changedReview{,} we presented the unfitted discontinuous Galerkin finite element method (UDG-FEM) for solving the EEG forward problem. In Section \ref{sec:theory}, we provided a mathematical formulation of the underlying model and in Section \ref{sec:methods}
a description of the overall method along with information regarding its implementation.

In our \changedReview{verification} and evaluation studies using multi-layer sphere models in Section \ref{subsec:sphereresults}, the results first of all show convergence of the UDG-FEM numerical solutions towards the analytical solutions when the mesh resolution is increased and, thereby, a better geometric approximation of the sphere model is achieved (\figurename{} \ref{fig:convergence}).
Furthermore, we showed that the accuracy of a discontinuous Galerkin FEM (DG-FEM) approach on a conforming mesh with tetrahedral elements could be met by our new UDG-FEM approach in models with a comparable number of degrees of freedom (DOFs) (\figurename{} \ref{fig:dgtetraudg}). Most importantly, an even better accuracy was achieved by the UDG-FEM when compared to DG-FEM on a structured mesh with hexahedral elements, even though the UDG-FEM mesh had a factor of about 79 times less DOFs than the DG-FEM mesh (\figurename{} \ref{fig:dghexaudg}).
 
The comparison with the DG-FEM approach in \figurename{} \ref{fig:dghexaudg} indicated the importance of the geometric representation of the domain boundary. This is in accordance to the findings in \cite{Engwer2015}, where a first DG-FEM approach
was presented for the EEG forward problem based on the subtraction approach \cite{Wolters2007b} and where the new approach was \changedReview{verified} in multi-layer sphere models on structured meshes with hexahedral elements. In contrast to Lagrange or Continuous Galerkin FEM (CG-FEM) approaches \cite{CHW:Wei2000,Schimpf2002,CHW:Gen2004,Wolters2007c,Lew2009b,Vallaghe2010,Pursiainen2011,Vorwerk2012,CHW:Med2015}, conservation properties could be proven for the new DG-FEM \cite{Engwer2015}. As the DG-FEM formulation is part of the UDG-FEM approach, the latter directly inherits the conservation properties of the former.  Furthermore,  in \cite{Engwer2015}, the geometric representation of the domain boundary could be identified as a main source of error for the numerical scheme.
Here, we reproduced this DG-FEM result in \figurename{} \ref{fig:dghexaudg} and showed, that even on a very coarse mesh with about 79 times less DOFs, the UDG-FEM approach achieved a similar accuracy when compared to DG-FEM.
Therefore, when taking the convergence results of \figurename{} \ref{fig:convergence} into account, it gets clear that on finer meshes UDG-FEM outperforms DG-FEM on structured meshes with hexahedral elements.

In \cite{Vallaghe2010}, a trilinear immersed finite element method (TI-FEM) has been proposed to solve the EEG forward problem. Similar to the UDG-FEM approach, it uses a level set function to represent the geometry and the basis functions are defined on a structured mesh. Since it is based on a continuous Galerkin approach, for the context of this paper, it is now abbreviated as TI-CG-FEM. Like UDG-FEM, also TI-CG-FEM modifies the local basis functions to incorporate the geometric information, but in contrast to UDG-FEM and like all other CG-FEM approaches, it does not fulfill the conservation property of electric charge. An additional important difference is that the TI-CG-FEM approach does not change the DOFs when compared with its corresponding CG-FEM approach, while UDG-FEM leads to an increase in DOFs when compared to its corresponding DG-FEM approach. In future studies, besides investigating the implications of fulfilling or not the conservation property and increasing or not the number of DOFs, it might furthermore be interesting to compare UDG-FEM and TI-CG-FEM with regard to thin skull compartments and limited FEM resolutions. This was done for DG-FEM and CG-FEM in \cite{Engwer2015}, where skull leakage problems, appearing in CG-FEM, could be strongly alleviated with DG-FEM. Such scenarios might get relevant, e.g., in infant studies \cite{Roche-Labarbe2008} or for temporal bone areas, where skull thickness is 2 mm or even less \cite[Table 2]{Kwon2006}.
When considering the comparison CG-FEM and TI-CG-FEM on the one hand and DG-FEM and UDG-FEM on the other, we find similar results with regard to accuracies. Both, TI-CG-FEM and UDG-FEM show better accuracies than their counterparts on structured hexahedral meshes. Furthermore, when compared to the results on conforming tetrahedral meshes, the accuracies are met by both approaches.
A more direct comparison is, however, difficult, since the parametrization of the multi-layer sphere model for \changedReview{verifying} TI-CG-FEM in \cite{Vallaghe2010} is different from ours with regard to number of compartments (4 here versus 3 in \cite{Vallaghe2010}), radii of the compartments, chosen conductivities and source eccentricities (in \cite{Vallaghe2010}, dipoles where considered along a single axis up to 2 mm to the next conductivity jump, i.e., in the terms used here, up to an eccentricity of 0.977).

Another method to improve the geometric representation of the standard CG-FEM approach on a structured hexahedral mesh has been presented in \cite{Wolters2007}.
In order to reduce the effect of a staircase approximation of the different compartment surfaces, the position of the mesh nodes is geometrically adapted based on the different conductivities in the surrounding elements.
The nodes are shifted towards the centroids of their minority elements by a fixed global shift length.
This length has to be chosen such that the determinant of the Jacobian of the element transformation remains positive.
As nodes are moved solely based on the voxel segmentation, the topology of the mesh is restricted to the one given by the segmentation and no further geometric information is included.
In \cite{Wolters2007}, the  geometry-adapted hexahedral CG-FEM method is \changedReview{verified} in a 4-layer sphere model, parameterized like here with the only exception that skull conductivity was chosen 0.0042 S/m (note, the lower the skull conductivity, the higher the expected numerical errors). A 2 mm mesh resolution was used and sources were modeled with subtraction and Saint-Venant approaches. 
Source positions where chosen along a single axis up to 2mm to the next conductivity jump (eccentricity of 0.974), so we compare the results with the median values of the 2 mm UDG-FEM model.
As we used the partial integration approach in the \changedReview{verification} of UDG-FEM, we compare the results with the Saint-Venant approach, which is methodologically closer than the subtraction approach.
With respect to the $\rdmp$, both methods perform reasonably well, with slightly better accuracy for the UDG-FEM approach.
The difference of the $\magp$ error is more pronounced, especially for radial sources.
For the geometry adapted CG-FEM, the $\magp$ error for radial sources is approximately 3.5 - 4 \%, while it is at approximately -1 \% (at eccentricity 0.9686) for the UDG-FEM approach.
As presented in \cite{Engwer2015}, the CG-FEM method with hexahedral elements might suffer from skull leakage currents on coarser meshes. The most important difference might thus be that this can be alleviated by employing the DG-FEM approach \cite{Engwer2015} and especially the here presented UDG-FEM, but not by the CG-FEM geometry-adaptation of \cite{Wolters2007}.

A different approach to solving the EEG forward problem is the boundary element method (BEM) \cite{Mosher1999,Acar2010,Gramfort2011,Stenroos2012}.
This method uses surface meshes and only takes the interfaces between different homogenized tissue compartments into account.
In principle, almost arbitrary partitioning of the domain into different tissue compartments is possible \cite{Kybic2006}.
In practice, however, mainly nested compartments with approximated holes are implemented, as the head model generation process is a difficult problem (see, e.g., \cite{Roche-Labarbe2008}).
For the UDG-FEM method, no such restrictions apply to the level set functions representing the different tissue boundaries.
UDG-FEM is especially suited for such complex shaped domains as presented here for source analysis and, for other applications, in \cite{Engwer2009}.
Although it was not directly part of this study, UDG-FEM can, like all FEM approaches, handle realistic brain tissue anisotropy, 
which is not the case for BEM.

We did not evaluate the computational load of the UDG-FEM approach in this paper.
When compared to the DG-FEM approach on a mesh with the same resolution, the UDG-FEM approach introduces more local basis functions on cells which are cut by a level set.
This increases the computational load as more DOFs have to be considered.
However, the additional DOFs are well invested in areas of special geometrical significance and as the UDG-FEM approach can achieve higher accuracies even on coarser meshes (see \figurename{}\ref{fig:dghexaudg}), the overall mesh resolution can thus even be strongly reduced. As a consequence, it then leads to a drastic reduction of the computational effort.
Another source of increased computational load can be found in the assembly of the linear system.
As the integration over cut cells introduces a subtriangulation, more quadrature points are generated and the bilinear form has to be evaluated more often.
However, since for the FEM transfer matrix approach, the linear system has to be assembled only once and then solved for each electrode \cite{Mun2012}, the resulting slightly increased overall setup time is rather negligible.
\changedReview{
  In a subsequent study, an evaluation of different solution techniques for UDG-FEM, including multigrid methods and a thorough investigation of the performance compared to competitive methods should be carried out.
  In this study, we only performed an example simulation in a realistically shaped head model.
  Due to the promising results in the multi-layer sphere models, a thorough evaluation of UDG-FEM in different realistic scenarios, especially with respect to the above mentioned limitations of conforming tetrahedral and hexahedral meshes, will be carried out in a subsequent study.
  In addition, an investigation of the volumetric error distribution in selected dipole and volume conductor configurations in sphere models for the different methods, including CG-FEM and TI-CG-FEM, using the analytic solution \cite{Munck1993} would provide further insight into the different model properties and will be subject of our future work.
}

\section{Conclusion}
\label{sec:conclusion}
We presented theory, \changedReview{verification} and evaluation of a new unfitted discontinuous Galerkin finite element method (UDG-FEM) for solving the EEG forward problem. In a multi-layer sphere model, the accuracy shows proper convergence behavior with increasing mesh resolution. When compared to the discontinuous Galerkin FEM (DG-FEM) discretization on a conforming tetrahedral mesh, the method performs similarly or more accurately. On a structured mesh with hexahedral elements, the UDG-FEM approach outperforms the DG-FEM approach with respect to accuracy. The EEG forward simulation in a realistic head model for an auditory source resulted in a smooth potential distribution over the head surface which is in line with practical findings for auditory evoked potentials.

\section*{Acknowlegment}
This work was partially supported by the Cluster of Excellence 1003 of the Deutsche Forschungsgemeinschaft (DFG EXC 1003 Cells in Motion), by EU project ChildBrain (Marie Curie Innovative Training Networks, grant agreement no. 641652)
and by the Priority Program 1665 of the Deutsche Forschungsgemeinschaft (DFG) (WO1425/5-1).

\bibliographystyle{IEEEtran}
\bibliography{literature}

\begin{thebibliography}{10}
\providecommand{\url}[1]{#1}
\csname url@samestyle\endcsname
\providecommand{\newblock}{\relax}
\providecommand{\bibinfo}[2]{#2}
\providecommand{\BIBentrySTDinterwordspacing}{\spaceskip=0pt\relax}
\providecommand{\BIBentryALTinterwordstretchfactor}{4}
\providecommand{\BIBentryALTinterwordspacing}{\spaceskip=\fontdimen2\font plus
\BIBentryALTinterwordstretchfactor\fontdimen3\font minus
  \fontdimen4\font\relax}
\providecommand{\BIBforeignlanguage}[2]{{%
\expandafter\ifx\csname l@#1\endcsname\relax
\typeout{** WARNING: IEEEtran.bst: No hyphenation pattern has been}%
\typeout{** loaded for the language `#1'. Using the pattern for}%
\typeout{** the default language instead.}%
\else
\language=\csname l@#1\endcsname
\fi
#2}}
\providecommand{\BIBdecl}{\relax}
\BIBdecl

\bibitem{Haemaelaeinen1993}
M.~H{\"a}m{\"a}l{\"a}inen \emph{et~al.}, ``Magnetoencephalography—theory,
  instrumentation, and applications to noninvasive studies of the working human
  brain,'' \emph{Reviews of Modern Physics}, vol.~65, no.~2, p. 413, 1993.

\bibitem{Mun2012}
\BIBentryALTinterwordspacing
J.~de~Munck \emph{et~al.}, \emph{EEG and MEG -- forward modelling. In
  \emph{R.Brette \& A.Destexhe (eds.): Handbook of Neural Activity
  Measurement.}}\hskip 1em plus 0.5em minus 0.4em\relax Cambridge University
  Press, ISBN 978-0-521-51622-8, 2012, chapter doi:
  10.1017/CBO9780511979958.006. [Online]. Available:
  \url{http://www.di.ens.fr/~brette/HandbookMeasurement/}
\BIBentrySTDinterwordspacing

\bibitem{Munck1993}
J.~de~{Munck} and M.~J. {Peters}, ``A fast method to compute the potential in
  the multisphere model,'' \emph{IEEE Trans.Biomed.Eng.}, vol.~40, no.~11, pp.
  1166--1174, 1993.

\bibitem{Mosher1999}
J.~C. Mosher \emph{et~al.}, ``{EEG} and {MEG}: forward solutions for inverse
  methods,'' \emph{IEEE Trans.Biomed.Eng.}, vol.~46, no.~3, pp. 245--259, 1999.

\bibitem{Acar2010}
Z.~A. Acar and S.~Makeig, ``Neuroelectromagnetic forward head modeling
  toolbox,'' \emph{J. Neurosci. Meth.}, vol. 190, no.~2, pp. 258--270, 2010.

\bibitem{Gramfort2011}
A.~Gramfort \emph{et~al.}, ``Forward field computation with {OpenMEEG},''
  \emph{Comp.Intell.Neurosci.}, vol. 2011, 2011.

\bibitem{Stenroos2012}
M.~Stenroos and J.~Sarvas, ``Bioelectromagnetic forward problem: isolated
  source approach revis (it) ed,'' \emph{Phys. Med. Biol.}, vol.~57, no.~11, p.
  3517, 2012.

\bibitem{Cook2006}
M.~J. Cook and Z.~J. Koles, ``A high-resolution anisotropic finite-volume head
  model for {EEG} source analysis,'' in \emph{Engineering in Medicine and
  Biology Society, 2006. EMBS'06. 28th Annual International Conference of the
  IEEE}.\hskip 1em plus 0.5em minus 0.4em\relax IEEE, 2006, pp. 4536--4539.

\bibitem{CHW:Wen2008}
K.~Wendel \emph{et~al.}, ``The influence of {CSF} on {EEG} sensitivity
  distributions of multilayered head models.'' \emph{IEEE Trans. Biomed. Eng.},
  vol.~55, no.~4, pp. 1454--1456, 2008.

\bibitem{Vatta2008}
F.~Vatta \emph{et~al.}, ``Solving the forward problem in {EEG} source analysis
  by spherical and fdm head modeling: a comparative analysis-biomed 2009.''
  \emph{Biomed.Sci.Instrument.}, vol.~45, pp. 382--388, 2008.

\bibitem{CHW:Mon2014}
V.~Montes-Restrepo \emph{et~al.}, ``Influence of skull modeling approaches on
  {EEG} source localization,'' \emph{Brain Topogr.}, vol.~27, pp. 95--111,
  2014.

\bibitem{CHW:Wei2000}
D.~Weinstein \emph{et~al.}, ``Lead-field bases for electroencephalography
  source imaging,'' \emph{Ann. Biomed.Eng.}, vol.~28, no.~9, pp. 1059--1066,
  2000.

\bibitem{Schimpf2002}
P.~H. Schimpf \emph{et~al.}, ``Dipole models for the {EEG} and {MEG},''
  \emph{IEEE Trans. Biomed. Eng.}, vol.~49, no.~5, pp. 409--418, 2002.

\bibitem{CHW:Gen2004}
N.~Gencer and C.~Acar, ``Sensitivity of {EEG} and {MEG} measurements to tissue
  conductivity.'' \emph{Phys.Med.Biol.}, vol.~49, pp. 701--717, 2004.

\bibitem{Wolters2007c}
C.~Wolters \emph{et~al.}, ``Numerical approaches for dipole modeling in finite
  element method based source analysis.'' \emph{Int.Cong.Ser.}, vol. 1300, pp.
  189--192, June 2007.

\bibitem{Lew2009b}
S.~Lew \emph{et~al.}, ``Improved {EEG} source analysis using low resolution
  conductivity estimation in a four-compartment finite element head model.''
  \emph{Human Brain Mapping}, vol.~30, no.~9, pp. 2862--2878, 2009.

\bibitem{Vallaghe2010}
S.~Vallagh{\'e} and T.~Papadopoulo, ``A trilinear immersed finite element
  method for solving the electroencephalography forward problem,'' \emph{SIAM
  J.Sci.Comp.}, vol.~32, no.~4, pp. 2379--2394, 2010.

\bibitem{Pursiainen2011}
S.~Pursiainen \emph{et~al.}, ``Forward simulation and inverse dipole
  localization with the lowest order {R}aviart-{T}homas elements for
  electroencephalography,'' \emph{Inverse Problems}, vol.~27, no.~4, p. 045003,
  2011.

\bibitem{Vorwerk2012}
J.~Vorwerk \emph{et~al.}, ``Comparison of boundary element and finite element
  approaches to the {EEG} forward problem,'' \emph{Biomed. Eng.}, vol.~57, no.
  SI-1, pp. 795--798, 2012.

\bibitem{CHW:Med2015}
T.~Medani \emph{et~al.}, ``{FEM} method for the {EEG} forward problem and
  improvement based on modification of the {S}aint {V}enant’s method,''
  \emph{Progress Electromag. Res.}, vol. 153, pp. 11--22, 2015.

\bibitem{Vorwerk2014}
J.~Vorwerk \emph{et~al.}, ``A guideline for head volume conductor modeling in
  {EEG} and {MEG},'' \emph{NeuroImage}, vol. 100, pp. 590--607, 2014.

\bibitem{CHW:Par2015}
E.~Paraskevopoulos \emph{et~al.}, ``Musical expertise is related to altered
  functional connectivity during audiovisual integration,'' \emph{PNAS}, vol.
  112, no.~40, pp. 12\,522--27, 2015.

\bibitem{CHW:Ayd2015}
{\"U}.~Aydin \emph{et~al.}, ``Combined {EEG/MEG} can outperform single modality
  {EEG} or {MEG} source reconstruction in presurgical epilepsy diagnosis,''
  \emph{PLoS ONE}, vol.~10, no.~3, p. e0118753, 2015.

\bibitem{Kwon2006}
J.-H. Kwon \emph{et~al.}, ``The thickness and texture of temporal bone in brain
  {CT} predict acoustic window failure of transcranial doppler.'' \emph{J.
  Neuroimag.}, vol.~16, pp. 347--52, 2006.

\bibitem{Sonntag2013}
H.~Sonntag \emph{et~al.}, ``Leakage effect in hexagonal {FEM} meshes of the
  {EEG} forward problem,'' \emph{Clinical EEG and Neuroscience}, p. P022, 2013.

\bibitem{Engwer2015}
\BIBentryALTinterwordspacing
C.~{Engwer} \emph{et~al.}, ``A discontinuous {G}alerkin method for the {EEG}
  forward problem,'' \emph{ArXiv e-prints}, Nov. 2015, submitted. [Online].
  Available: \url{arXiv:1511.04892}
\BIBentrySTDinterwordspacing

\bibitem{Dannhauer2011}
M.~Dannhauer \emph{et~al.}, ``Modeling of the human skull in {EEG} source
  analysis,'' \emph{Human Brain Mapping}, vol.~32, no.~9, pp. 1383--1399, 2011.

\bibitem{Wolters2007}
C.~H. Wolters \emph{et~al.}, ``Geometry-adapted hexahedral meshes improve
  accuracy of finite-element-method-based {EEG} source analysis.'' \emph{IEEE
  Trans. Biomed. Eng.}, vol.~54, no.~8, pp. 1446--1453, 2007.

\bibitem{Bastian2009}
P.~Bastian and C.~Engwer, ``An unfitted finite element method using
  discontinuous {G}alerkin,'' \emph{Int. J. Numer. Methods Eng.}, vol.~79,
  no.~12, pp. 1557--1576, 2009.

\bibitem{Heimann2013}
F.~Heimann \emph{et~al.}, ``An unfitted interior penalty discontinuous
  {G}alerkin method for incompressible {N}avier--{S}tokes two-phase flow,''
  \emph{Int. J. Numer. Meth. in Fluids}, vol.~71, no.~3, pp. 269--293, 2013.

\bibitem{Engwer2014}
C.~Engwer and S.~Westerheide, ``Heterogeneous coupling for implicitly described
  domains,'' in \emph{Domain Decomposition Methods in Science and Engineering
  XXI}.\hskip 1em plus 0.5em minus 0.4em\relax Springer, 2014, pp. 809--817.

\bibitem{Burman2015}
\BIBentryALTinterwordspacing
E.~Burman \emph{et~al.}, ``A {Cut} {Discontinuous} {Galerkin} {Method} for the
  {Laplace}-{Beltrami} {Operator},'' \emph{ArXiv e-prints}, Jul. 2015.
  [Online]. Available: \url{arXiv:1507.05835}
\BIBentrySTDinterwordspacing

\bibitem{Ern2008}
A.~Ern \emph{et~al.}, ``A discontinuous {G}alerkin method with weighted
  averages for advection--diffusion equations with locally small and
  anisotropic diffusivity,'' \emph{IMA J. Num. Anal.}, 2008.

\bibitem{Han2004}
X.~Han \emph{et~al.}, ``{\color{black}{CRUISE}: cortical reconstruction using
  implicit surface evolution},'' \emph{NeuroImage}, vol.~23, no.~3, pp.
  997--1012, 2004.

\bibitem{Papadopoulo2007}
T.~Papadopoulo and S.~Vallagh{\'e}, ``Implicit meshing for finite element
  methods using levelsets,'' in \emph{Computer Vision, 2007. ICCV 2007. IEEE
  11th International Conference on}.\hskip 1em plus 0.5em minus 0.4em\relax
  IEEE, 2007, pp. 1--8.

\bibitem{Engwer2016}
\BIBentryALTinterwordspacing
C.~Engwer and A.~N{\"u}{\ss}ing, ``Geometric integration over irregular domains
  with topologic guarantees,'' \emph{ArXiv e-prints}, Jan. 2016, submitted.
  [Online]. Available: \url{arXiv:1601.03597}
\BIBentrySTDinterwordspacing

\bibitem{Tanzer2005}
I.~O. Tanzer \emph{et~al.}, ``Representation of bioelectric current sources
  using {W}hitney elements in the finite element method,''
  \emph{Phys.Med.Biol.}, vol.~50, no.~13, p. 3023, 2005.

\bibitem{Wolters2007b}
C.~Wolters \emph{et~al.}, ``Numerical mathematics of the subtraction method for
  the modeling of a current dipole in {EEG} source reconstruction using finite
  element head models.'' \emph{SIAM J.Sci.Comp.}, vol.~30, no.~1, pp. 24--45,
  2007.

\bibitem{CHW:Vor2016}
J.~Vorwerk, ``New finite element methods to solve the {EEG/MEG} forward
  problem,'' {Dissertation}, Fachbereich Mathematik und Informatik,
  Westf{\"a}lische Wilhelms-Universit{\"a}t M{\"u}nster, Germany, February
  2016.

\bibitem{Vallaghe2010b}
S.~Vallagh{\'e} \emph{et~al.}, ``{\color{black}The adjoint method for general
  {EEG} and {MEG} sensor-based lead field equations},'' \emph{Physics in
  {M}edicine and {B}iology}, vol.~54, no.~1, p. 135, 2008.

\bibitem{Bastian2008a}
{P. Bastian \emph{et al.}}, ``A generic grid interface for parallel and
  adaptive scientific computing. {P}art {I}: {A}bstract framework,''
  \emph{Computing}, vol.~82, no. 2--3, pp. 103--119, 2008.

\bibitem{Bastian2008b}
{P. Bastian \emph{et al}}, ``A generic grid interface for parallel and adaptive
  scientific computing. {P}art {II}: {I}mplementation and tests in {DUNE},''
  \emph{Computing}, vol.~82, no. 2--3, pp. 121--138, 2008.

\bibitem{Blatt2007}
M.~Blatt and P.~Bastian, ``The iterative solver template library,'' in
  \emph{Applied Parallel Computing. State of the Art in Scientific Computing},
  ser. Lecture Notes in Computer Science, B.~K\r{a}gstr\"om, E.~Elmroth,
  J.~Dongarra, and J.~Wa\'sniewski, Eds., vol. 4699.\hskip 1em plus 0.5em minus
  0.4em\relax Springer, 2007, pp. 666--675.

\bibitem{Engwer2012}
C.~Engwer and F.~Heimann, ``{DUNE-UDG}: a cut-cell framework for unfitted
  discontinuous {G}alerkin methods,'' in \emph{Advances in {DUNE}}.\hskip 1em
  plus 0.5em minus 0.4em\relax Springer, 2012, pp. 89--100.

\bibitem{Dedner2014}
A.~Dedner \emph{et~al.}, ``The {DUNE-ALUGrid} module,'' \emph{arXiv preprint
  arXiv:1407.6954}, 2014.

\bibitem{Bastian2010}
P.~Bastian \emph{et~al.}, ``Generic implementation of finite element methods in
  the distributed and unified numerics environment ({DUNE}),''
  \emph{Kybernetika}, vol.~46, no.~2, pp. 294--315, 2010.

\bibitem{Bastian2012}
{P. Bastian \emph{et al.}}, ``Algebraic multigrid for discontinuous {G}alerkin
  discretizations of heterogeneous elliptic problems,'' \emph{Numerical Linear
  Algebra with Applications}, vol.~19, no.~2, pp. 367--388, 2012.

\bibitem{Davis2004}
T.~A. Davis, ``Algorithm 832: Umfpack v4.3---an unsymmetric-pattern
  multifrontal method,'' \emph{ACM Trans. Math. Softw.}, vol.~30, no.~2, pp.
  196--199, Jun. 2004.

\bibitem{Si2015}
H.~Si, ``Tetgen, a delaunay-based quality tetrahedral mesh generator,''
  \emph{ACM Trans. Math. Softw.}, vol.~41, no.~2, pp. 11:1--11:36, Feb. 2015.

\bibitem{CHW:Li2014}
G.~Li \emph{et~al.}, ``Measuring the dynamic longitudinal cortex development in
  infants by reconstruction of temporally consistent cortical surfaces,''
  \emph{NeuroImage}, vol.~90, pp. 266--279, 2014.

\bibitem{Vese2002}
L.~A. Vese and T.~F. Chan, ``A multiphase level set framework for image
  segmentation using the {M}umford and {S}hah model,'' \emph{Int.J.Comp.Vis.},
  vol.~50, no.~3, pp. 271--293, 2002.

\bibitem{Oostenveld2002}
R.~Oostenveld and T.~F. Oostendorp, ``Validating the boundary element method
  for forward and inverse {EEG} computations in the presence of a hole in the
  skull,'' \emph{Human Brain Mapping}, vol.~17, no.~3, pp. 179--192, 2002.

\bibitem{Roche-Labarbe2008}
Roche-Labarbe \emph{et~al.}, ``High-resolution electroencephalography and
  source localization in neonates,'' \emph{Human Brain Mapping}, vol.~29,
  no.~2, pp. 167--176, 2008.

\bibitem{Taubin1995}
G.~Taubin, ``A signal processing approach to fair surface design,'' in
  \emph{Proceedings of the 22nd annual conference on Computer graphics and
  interactive techniques}.\hskip 1em plus 0.5em minus 0.4em\relax ACM, 1995,
  pp. 351--358.

\bibitem{Kybic2006}
J.~Kybic \emph{et~al.}, ``Generalized head models for {MEG/EEG}: {B}oundary
  element method beyond nested volumes,'' \emph{Phys.Med.Biol.}, vol.~51,
  no.~5, p. 1333, 2006.

\bibitem{Engwer2009}
C.~Engwer, ``An unfitted discontinuous {G}alerkin scheme for micro-scale
  simulations and numerical upscaling,'' Ph.D. dissertation, Heidelberg
  University, 2009.

\end{thebibliography}

\end{document}